\documentclass[%
reprint,
superscriptaddress,
showpacs,preprintnumbers,
 amsmath,amssymb,
aps,
pre,
floatfix
]{revtex4-1}
\usepackage{amsmath}
\usepackage{graphicx}
\DeclareMathSizes{12}{17.28}{9}{7}

\usepackage{float}
\usepackage{dcolumn}
\usepackage{bm}
\usepackage{bbold}
\usepackage[multidot]{grffile}
\usepackage[colorlinks=true, breaklinks=true]{hyperref}
\usepackage[mathlines]{lineno}

\usepackage{xcolor}

\usepackage{xspace}

\usepackage[utf8]{inputenc}
\usepackage{tikz}
\usetikzlibrary{shapes.geometric,fit,arrows,babel}
\usepackage[utf8]{inputenc} 
\usepackage[T1]{fontenc}    
\usepackage{hyperref}
\hypersetup{colorlinks,
citecolor=blue
}
\usepackage{url}            
\usepackage{booktabs}       
\usepackage{amsfonts}       
\usepackage{nicefrac}       
\usepackage{microtype}      
\usepackage{lipsum}
\usepackage{amsmath,amssymb,amsthm}
\usepackage[normalem]{ulem}
\usepackage[colorinlistoftodos,prependcaption,textsize=tiny]{todonotes}
\usepackage{graphicx}
\usepackage{wrapfig}

\newcommand{\R}[1]{\textcolor{black}{{#1}}} 
\definecolor{Agreen}{rgb}{0.1, 0.6, 0.1} 

\usepackage{multirow}

\usepackage[linesnumbered,ruled,vlined]{algorithm2e}
\usepackage[noend]{algpseudocode}
\SetKwInput{KwInput}{Input}                
\SetKwInput{KwOutput}{Output}              

\SetCommentSty{mycommfont}

\newcommand{\marius}[1]{\textcolor{black}{#1}}

\begin{document}


\title{Combined effects of STDP and homeostatic structural plasticity on coherence resonance}

\author{Marius E. Yamakou}
\email{marius.yamakou@fau.de}
\affiliation{Department of Data Science, Friedrich-Alexander-Universit\"{a}t Erlangen-N\"{u}rnberg, Cauerstr. 11, 91058 Erlangen, Germany}
\affiliation{Max-Planck-Institut f\"ur Mathematik in den Naturwissenschaften, Inselstr. 22, 04103 Leipzig, Germany}

\author{Christian Kuehn}
\email{ckuehn@ma.tum.de}
\affiliation{Faculty of Mathematics, Technical University of Munich, Boltzmannstrasse 3, 85748 Garching bei München, Germany}
\affiliation{Complexity Science Hub Vienna, Josefstädter Strasse 39, 1080 Vienna, Austria}

\date{\today}

\begin{abstract}
Efficient processing and transfer of information in neurons have been linked to noise-induced resonance phenomena such as coherence resonance (CR), and adaptive rules in neural networks have been mostly linked to two prevalent mechanisms: spike-timing-dependent plasticity (STDP) and homeostatic structural plasticity (HSP). Thus, this paper investigates CR in small-world and random adaptive networks of Hodgkin-Huxley neurons driven by STDP and HSP. Our numerical study indicates that the degree of CR strongly depends, and in different ways, on the adjusting rate parameter $P$, which controls STDP, on the characteristic rewiring frequency parameter $F$, which controls HSP, and on the parameters of the network topology. In particular, we found two robust behaviors: (i) Decreasing  $P$ (which enhances the weakening effect of STDP on synaptic weights) and decreasing $F$ (which slows down the swapping rate of synapses between neurons) always leads to higher degrees of CR in small-world and random networks, provided that the synaptic time delay parameter $\tau_c$ has some appropriate values. (ii) Increasing the synaptic time delay $\tau_c$ induces multiple CR (MCR) \marius{--- the occurrence of multiple peaks in the degree of
coherence as $\tau_c$ changes ---} in small-world and random networks, with MCR becoming more pronounced at smaller values of $P$ and $F$. Our results imply that STDP and HSP can jointly play an essential role in enhancing the time precision of firing necessary for optimal information processing and transfer in neural systems and could thus have applications in designing networks of noisy artificial neural circuits engineered to use CR to optimize information processing and transfer.
\end{abstract}

\maketitle

    
\section{Introduction}\label{Sec. I}
Noise is an inherent part of many complex systems and has been observed to induce a wide variety of phenomena \cite{benzi1982stochastic,hou2005transfer,hanggi2002stochastic,horsthemke1984noise,mcdonnell2011benefits}. In the case of neural systems,  noise-induced resonance phenomena, such as stochastic resonance \cite{hanggi2002stochastic,longtin1993stochastic,gluckman1998stochastic,bulsara1991stochastic}, self-induced stochastic resonance (SISR) \cite{muratov2005self,yamakou2022levy,yamakou2022diversity,yamakou2017simple}, inverse stochastic resonance \cite{yamakou2017simple,gutkin2009inhibition,buchin2016inverse,yamakou2018weak} and coherence resonance (CR) \cite{pikovsky1997coherence,yamakou2022optimal,yamakou2019control,yamakou2023coherence}, have been shown to play important functional roles in information processing, including amongst others, the detection of weak input signals in neural networks \cite{vazquez2017stochastic}, and the local optimal information transfer between the input and output spike in neurons \cite{buchin2016inverse}. 

Due to the importance of these noise-induced resonance phenomena in neural information processing and transfer, the
research on their dynamics in single neurons and neural networks with different levels of complexity has been extensively investigated over the last decades. The literature on the topic is abundant, including: \marius{the effects of different types of noise \cite{sun2008spatial,yamakou2022levy,lu2020inverse},
network size and topology \cite{gao2001stochastic,guo2009stochastic,liu2014effects,gosak2010optimal,toral2003system,semenova2018weak},
various spatial distributions on resonance \cite{tuckwell2011effects},
electrical synaptic couplings \cite{uzuntarla2017inverse,yamakou2022optimal}, 
inhibitory or excitatory chemical synaptic couplings \cite{uzuntarla2017inverse}, 
hybrid of electrical and chemical synaptic couplings \cite{yilmaz2013stochastic}, 
time delay in the synaptic couplings \cite{wang2009delay,liu2014effects}, and time-varying strength of synaptic couplings \cite{yilmaz2015enhancement}.
The interplay between different noise-induced resonance phenomena has also been investigated; e.g., the interplay between CR and SISR in multiplex neural networks \cite{yamakou2022optimal,yamakou2019control} and the interplay between SR and ISR \cite{zamani2020concomitance} in a basic neural circuit.
The control of stochastic resonance in experiments \cite{gammaitoni1999controlling,gluckman1996stochastic} and the resonance induced by chaos has also been extensively studied \cite{sinha1998deterministic,zambrano2007chaos,sinha1999noise,nobukawa2019controlling}.}

However, to this day, insufficient attention has been devoted to the behavior of these noise-induced resonance phenomena in adaptive neural networks. The effects of the inherently adaptive nature of neural networks on these noise-induced phenomena and hence on neural information processing and transfer are important and cannot be overlooked, especially when a deep understanding of neural dynamics is required. Adaptation in neural networks precisely refers to the ability to change the strength of the synaptic couplings over time and/or the architecture of the neural network topology via some specific rules. Adaptive rules in neural networks have been mostly linked to two important mechanisms: (i) spike-timing-dependent plasticity (STDP) and (ii) structural plasticity (SP). 

STDP describes how the synaptic weights get modified by repeated pairings of pre-and post-synaptic membrane potentials with the sign and the degree of the modification dependent on the relative timing of the neurons firing. Depending on the precise timing of the pre-and post-synaptic action potentials, the synaptic weights can exhibit either long-term potentiation (LTP) --- persistent strengthening of synapses --- or long-term depression (LTD) --- persistent weakening of synapses \cite{gerstner1996neuronal,markram1997regulation,morrison2007spike}. 

On the other hand, SP describes the mechanisms that rewire connectivity between neurons over time by altering (creating, pruning, or swapping) synaptic connections between neurons, thereby changing the architecture of the network or the brain as a whole. The initial evidence for SP came from histological studies of spine density following new sensory experience or training~\cite{greenough1988anatomy}. Today, there is good evidence that the micro-connectome, which describes the connectome at the level of individual synapses, rewires~\cite{van2017rewiring,bennett2018rewiring}.
 Moreover, experiments suggest that the rewiring of cortical circuits contributes to the reorganization of neural firing during developmental experience-dependent plasticity in the visual cortex~\cite{ko2013emergence}, and the rewiring involves changing either the number of synapses or the location of those synapses~\cite{holtmaat2009experience}.

The question of brain networks adhering to specific architectures, including small-world and random networks, despite their time-varying dynamics was recently considered \cite{hilgetag2016brain,valencia2008dynamic}. Homeostasis is the ability of a system to maintain a relatively stable state that persists despite changes in the system's evolution. It is argued that SP must be accompanied by some homeostatic mechanisms that prevent neural networks from losing the architectures required for optimal information processing and transfer \cite{turrigiano2012homeostatic}. It has been shown that small-world and random networks can benefit from homeostasis via an increase in the efficiency of information processing \cite{butz2014homeostatic}. Thus, in the present study, we focus on small-world and random networks driven by homeostatic structural plasticity (HSP), i.e., time-varying small-world and random networks that adhere to their respective topology over time.

Combining STDP and other forms of homeostatic plasticity rules, such as synaptic scaling \cite{pozo2010unraveling}, has provided knowledge on how to keep neural network dynamics in a stable and biologically plausible regime \cite{watt2010homeostatic} and to support non-trivial computations underlying many optimization tasks \cite{galtier2013biological}. Motivated by these studies, we focus, in the present paper, on one noise-induced resonance phenomenon, namely, CR in adaptive neural networks driven by both STDP and HSP. 

\marius{CR occurs when the regularity of noise-induced oscillations of an excitable system without a periodic input signal is a non-monotonic function of the noise amplitude, or another system parameter \cite{pikovsky1997coherence,gosak2010optimal}. Important cases of CR occur when the system remains sufficiently close to a Hopf bifurcation 
\cite{pikovsky1997coherence,yamakou2019control}, or a saddle-node bifurcation of limit cycles \cite{jia2011coherence}.} As we pointed out earlier, most of the previous literature on CR reports that the behavior of this noise-induced phenomenon is in non-adaptive neural networks. The studies investigating CR in adaptive neural networks have considered only STDP. For example, in \cite{yu2015spike}, it is found that CR depends significantly on the features of STDP and connectivity in Newman–Watts small-world neural networks. It is also demonstrated that the network structure plays a constructive role in the degree of CR: the degree of CR reaches a peak when the number of random shortcuts is intermediate. Moreover, as the adjusting rate parameter of the STDP increases, the average coupling strength of the network is weakened. Consequently, the degree of CR largely deteriorates. The same study also shows that more connections are needed to optimize the temporal coherence-related random shortcuts with a larger adjusting rate of the STDP. 

In \cite{xie2016effect}, it is found that in a Newman–Watts small-world neural network, multiple CR induced by the time delay of the STDP can be either enhanced or suppressed as the adjusting rate of STDP changes, depending on the number of added random shortcuts is the network. 
In \cite{xie2018spike}, it is found that in a scale-free neural network with autaptic time delay, as the adjusting rate parameter of STDP increases, multiple coherence resonance enhances and becomes strongest at an intermediate value of this adjusting rate parameter, indicating that there is optimal STDP that can most strongly enhance the multiple CR induced by \marius{time delay}.

\marius{To the best of our knowledge, no previous study has considered CR in time-varying neural networks driven by STDP and HSP. Thus, the overarching scientific question of this work is to determine whether and, if so, how can STDP and HSP jointly affect the degree of CR in small-world and random neural networks. To address this main question, we investigate the following questions in small-world and random networks:
(i) How do the adjusting rate of the STDP rule and the characteristic rewiring frequency of the HSP rule jointly affect the degree of CR? (ii) How do the synaptic time delay, the rewiring frequency of the HSP rule, and the adjusting rate of the STDP rule jointly affect the degree of CR? (iii) How do the average degree of network connectivity, the rewiring frequency of the HSP rule, and the adjusting rate of the STDP rule jointly affect the degree of CR? (iv) How do the rewiring probability of the Watts-Strogatz small-world network, the rewiring frequency of the HSP rule, and the adjusting rate of the STDP rule jointly affect the degree of CR? We address these questions using extensive numerical simulations.}

The numerical results indicate that the degree of CR depends in one general way on parameters controlling STDP and HSP, and in various ways, on the interval in which network topology parameters lie. For example, it is found that decreasing the STDP and HSP parameters ($P$ and $F$, respectively) always leads to higher degrees of CR in small-world and random networks, provided that the synaptic time delay parameter $\tau_c$ is fixed to some suitable values. Moreover, increasing the synaptic time delay $\tau_c$ induces multiple CR (MCR) in small-world and random networks, with MCR becoming more pronounced as the STDP and HSP parameters decrease. On the other hand, the degree of CR is found to vary non-monotonically when the network parameters, including the rewiring probability, average degree, and synaptic time delay, vary.

The rest of the paper is organized as follows: In Sec. \ref{Sec. II}, we describe the mathematical neural network model, the STDP learning rule, and the strategy of HSP that would allow the time-varying small-world and random networks to adhere to their respective architecture. In Sec. \ref{Sec. III}, we describe the computational methods used. In Sec. \ref{Sec. IV}, we present and discuss numerical results. We summarize and conclude in Sec. \ref{Sec. V}.

\section{Model description}\label{Sec. II}
\subsection{Neural Network Model}
\marius{Unlike the mathematically simpler but biophysically less relevant neuron models (e.g., the FitzHugh-Nagumo model \cite{fitzhugh1961impulses}), the more complicated Hodgkin-Huxley (HH) \cite{hodgkin1952quantitative} neuron model can provide experimentally testable hypotheses that are mature enough to guide experiments in \textit{vivo} and \textit{vitro}.}
Thus, as a paradigmatic model with well-known biophysical relevance, we study the effects of STDP and HSP in a network of HH neurons, whose membrane potential is governed by
\begin{eqnarray}\label{eq:1}
\begin{split}
\hspace{-6.0mm}\displaystyle{C_m\frac{dV_{i}}{dt}} = &- g_{_{\text{Na}}}^{max}m_{i}^3h_{i}(V_{i}-V_{_{\text{Na}}}) - g_{_{\text{K}}}^{max}n_{i}^4(V_{i}-V_{_{\text{K}}})\\ 
&- g_{_{\text{L}}}^{max}(V_{i}-V_{_{\text{L}}}) + I_{i}^{syn}(t),
\end{split}
\end{eqnarray}
where the variable $V_i$, $i=1,...,N,$ represents the membrane potential (measured in $\mathrm{mV}$) of neuron $i$, and $t$ is the time (measured in $\mathrm{ms}$).
The capacitance of the membrane of each neuron is represented by $C_m = 1 \mathrm{\mu F/cm^3}$.  The conductances $g_{_{\text{Na}}}^{max}=120$ $\mathrm{mS/cm^2}$, $g_{_{\text{K}}}^{max}=36$ $\mathrm{mS/cm^2}$,  and $g_{_{\text{L}}}^{max}=0.3$ $\mathrm{mS/cm^2}$ 
respectively denote the maximal sodium, potassium, and leakage conductance when all ion channels are open. The potentials $V_{_{\text{Na}}}= 50.0$ $\mathrm{mV}$, $V_{_{\text{K}}} = - 77.0$ $\mathrm{mV}$, and $V_{_{\text{L}}}=-54.4$ $\mathrm{mV}$ are the reversal potentials for sodium,
potassium and leak channels, respectively. 

$m_i^3h_i$ and $n_i^4$ in Eq.~\eqref{eq:1} are respectively the mean portions of the open sodium and potassium ion channels within a membrane patch, while
$x_{i}=\{m_{i},h_{i},n_{i}\}$ represent auxiliary dimensionless $[0,1]$-valued dynamical variables representing 
the sodium activation, sodium inactivation, and potassium activation, respectively. The dynamics of these gating variables $(x_i = m_i,h_i,n_i)$, depending on the voltage-dependent opening and closing rate functions $\alpha_{{x_{i}}}(V_i)$ and $\beta_{{x_{i}}}(V_i)$, are given by
\begin{equation}\label{eq:2}
 \displaystyle{\frac{dx_{i}}{dt}} = \alpha_{{x_{i}}}(V_{i})(1 - x_{i}) - \beta_{{x_{i}}}(V_{i})x_{i} + \displaystyle{\xi_{x_i}(t)},
\end{equation}
where the rate functions are given by
\begin{equation}\label{eq:3}
\begin{split}
\left\{\begin{array}{lcl}
\alpha_{m_i}(V_i) &=& \displaystyle{ \frac{(V_i+40)/10}{1-\exp{[-(V_i+40)/10}]},}\\ [4.0mm]
\beta_{m_i}(V_i) &=&  \displaystyle{4\exp{\big[-(V_i+65)/18\big]},}\\[4.0mm] 
\alpha_{h_i}(V_i) &=&  \displaystyle{0.07\exp{\big[-(V_i+65)/20\big]},}\\[2.0mm] 
\beta_{h_i}(V_i) &=&  \displaystyle{\frac{1}{1+\exp{[-(V_i+35)/10]}},}\\[4.0mm] 
\alpha_{n_i}(V_i) &=&  \displaystyle{\frac{(V_i+55)/100}{1-\exp{[-(V_i+55)/10}]},}\\[4.0mm]
\beta_{n_i}(V_i) &=&  \displaystyle{0.125\exp{\big[-(V_i+65)/80\big]}}.
\end{array}\right.
\end{split}
\end{equation}

$\xi_{x_i}(t)$ $(x_{i}=\{m_{i},h_{i},n_{i}\})$ in Eq.\eqref{eq:2} represent ion channel noises in the HH neurons. We will use the sub-unit noise as the ion channel noises \cite{fox1997stochastic,white2000channel} where $\xi_{x_i}(t)$ $(x_{i}=\{m_{i},h_{i},n_{i}\})$ are given by independent zero mean Gaussian white noise sources whose autocorrelation functions are given as
\begin{equation}\label{eq:4}
\begin{split}
\left\{\begin{array}{lcl}
 \langle\xi_{m_i}(t)\xi_{m_i}(t')\rangle = \displaystyle{ \frac{2\alpha_{m_i}(V_i)\beta_{m_i}(V_i)}{\rho_{Na}A_i\big[\alpha_{m_i}(V_i)+\beta_{m_i}(V_i)\big]}\delta(t-t')},\\ [5.0mm]
\langle\xi_{h_i}(t)\xi_{h_i}(t')\rangle = \displaystyle{ \frac{2\alpha_{h_i}(V_i)\beta_{h_i}(V_i)}{\rho_{Na}A_i\big[\alpha_{h_i}(V_i)+\beta_{h_i}(V_i)\big]}\delta(t-t')},\\ [5.0mm]
\langle\xi_{n_i}(t)\xi_{n_i}(t')\rangle = \displaystyle{ \frac{2\alpha_{n_i}(V_i)\beta_{n_i}(V_i)}{ \rho_KA_i\big[\alpha_{n_i}(V_i)+\beta_{n_i}(V_i)\big]}\delta(t-t')},
\end{array}\right.
\end{split}
\end{equation}
where $\rho_{Na}$ and $\rho_K$ are the sodium and potassium channel densities, respectively, and $A_i$ is the membrane patch area (\marius{measured in $\mathrm{\mu m^2}$}) of the $i$th neuron. \R{It is worth noting that the membrane patch area in a neuron refers to the area of the cell membrane that is being studied or measured. It is not necessarily the same as the total surface area of the neuron or the soma (cell body). Neuroscientists often use patch-clamp electrophysiology to study the electrical properties of neurons. This technique involves attaching a small glass pipette to the surface of a neuron and creating a patch of the membrane that is then used to measure the neuron's electrical activity. The patch size can vary depending on the experimental setup, and researchers can choose to study membrane patches from different parts of the neuron, including the soma, dendrites, and axon. So, the membrane patch area refers specifically to the area of the cell membrane that is being studied, and it may or may not be the same as the total surface area of the neuron or soma.} For simplicity, we assume that all the neurons in the network have the same membrane patch area, i.e., we choose $A_1=A_2=...=A_N=A$.

\subsection{Synapses and STDP Rule}
The term $I_{i}^{syn}(t)$ in Eq.\eqref{eq:1} models the inhibitory and uni-directional chemical synapses between the neurons and also controls the STDP learning rule between these connected neurons. The synaptic current $I_{i}^{syn}(t)$ of the $i$th neuron at time $t$ is given by
\begin{equation}\label{eq:5}
I_{i}^{syn}(t) = - \sum_{j=1(\neq i)}^{N}\ell_{ij}(t)g_{ij}(t)s_j(t)\big[V_i(t)- V_{syn}\big],
\end{equation}
where the synaptic connectivity matrix $L (=\{\ell_{ij}(t)\})$ has $\ell_{ij}(t)=1$ if the neuron $j$ is pre-synaptic to the neuron $i$; otherwise, $\ell_{ij}(t)=0$. The synaptic connection is modeled as a time-varying small-world network or a time-varying random network.
The small-world and the random network are generated using the Watts-Strogatz algorithm \cite{watts1998collective}, where for a given average degree $\langle k \rangle$, the value of the rewiring probability $\beta\in[0,1]$ in the algorithm will determine whether we generate a small-world network (i.e., when $\beta\in(0,1)$) or a completely random network (i.e., when $\beta=1$). This study does not consider regular networks (i.e., when $\beta=0$). The control parameters of the network topology are the average degree $\langle k \rangle$ and the rewiring probability $\beta$.

The fraction of open synaptic ion channels at time $t$ of the $j$th neuron is represented by $s_j(t)$ in Eq.\eqref{eq:5} and its time-evolution is governed by \cite{yu2015spike}:
 \begin{equation}\label{eq:5a}
\frac{ds_j}{dt} = \frac{2(1 - s_j)}{1 + \displaystyle{\exp\Bigg[- \frac{V_j(t-\tau_c)}{V_{shp}}\Bigg]}}-s_j,
 \end{equation}
 where $V_j(t-\tau_c)$ is the action potential of the pre-synaptic neuron $j$ at earlier time $t-\tau_c$,  $\tau_c$ (in the unit of $\mathrm{ms}$) is delayed time which will be used as a control parameter of the chemical synapses. $V_{shp}=5.0$ $\mathrm{mV}$ determines the threshold of the membrane potential above which the post-synaptic neuron $i$ is affected by the pre-synaptic neuron $j$. 
 
The weight of the synaptic connection from the pre-synaptic $j$th neuron to the post-synaptic $i$th neuron is represented by $g_{ij}(t)$ in Eq.\eqref{eq:5}.  According to the STDP mechanism, with increasing time $t$, the synaptic strength $g_{ij}$ for each synapse is updated with a nearest-spike pair-based STDP rule \cite{morrison2007spike}. The synaptic coupling strength $g_{ij}(t)$ update via the synaptic modification function $M$, which depends on the current value of the synaptic weight $g_{ij}(t)$ and which is defined as follows \cite{xie2018spike}:
  \begin{eqnarray}\label{eq:6}
\left\{\begin{array}{lcl}
 g_{ij}(t + \Delta t) = g_{ij}(t) + \Delta g_{ij},\\[3.0mm]
\Delta g_{ij}=g_{ij}(t)M(\Delta t),
  \\[3.0mm]
M(\Delta t)=
  \left\{
\begin{array}{ll}
\displaystyle{P\exp{(-\lvert\Delta t\rvert/\tau_{p})}\:\:\text{if}~\Delta t>0 }\\[1.0mm]
\displaystyle{- D\exp{(-\lvert\Delta t\rvert/\tau_{d})}\:\:\text{if}~\Delta t<0}\\[1.0mm]
0 \:\:\text{if}~\Delta t=0,
\end{array} 
\right.
\end{array}\right.
\end{eqnarray}
where $\Delta t=t_i -t_j$, $t_i$ (or $t_j$) represents the spiking time of the $i$th ($j$th) neuron. The amount of synaptic modification is controlled by the adjusting potentiation and depression rate parameters of the STDP rule, represented by $P$ and $D$, respectively. \marius{The potentiation and depression temporal windows of the synaptic modification are controlled by $\tau_p$ and $\tau_d$, respectively.}

Experimental studies have demonstrated that the temporal window for synaptic weakening is approximately the same as that for synaptic strengthening \cite{bi1998synaptic,feldman2005map,song2000competitive}. Synaptic potentiation is consistently induced when the post-synaptic spike generates within a time window of $20$ $\mathrm{ms}$ after the pre-synaptic spike, and depression is caused conversely. Furthermore, STDP is usually viewed as depression-dominated. Thus, in this study, we set the temporal window parameters at $\tau_p = \tau_d$ = 20 $\mathrm{ms}$ \cite{song2000competitive} and $D/P=1.05$, and we choose $P$ as the control parameter of the STDP rule.

\subsection{Time-varying Networks and HSP Rule}
Here, we consider  the  network to have a small-world structure \cite{bassett2006small,liao2017small,bassett2006adaptive,muldoon2016small}, constructed by the Watts-Strogatz network algorithm \cite{watts1998collective}, whose Laplacian matrix is a zero row sum matrix with average degree $\langle k\rangle$ and rewiring probability $\beta\in(0,1)$. To study the effects of HSP, i.e., the effects of time-dependence of the network topology such that it adheres to a small-world topology as time advances, we implement the following strategy for the time-evolution of synapses (edges): \textit{at each integration
time step $dt$, if there is an edge between two distant neighbors, it is rewired to a nearest neighbor of one of the neurons (nodes) with probability $(1 - \beta)Fdt$. If the edge is between two nearest neighbors, then with probability $\beta Fdt$, it is replaced by a connection to a randomly chosen distant node}.  
In case of a random network (i.e., when $\beta=1$ in Watts-Strogatz network algorithm), we implement the following strategy for the time-evolution of edges: \textit{at each integration
time step $dt$, if there is an edge between node $i$ and $j$, it will be rewired such that node $i$ ($j$) connects to any another node excluding $j$ ($i$) with probability $\big(1-\frac{\langle k\rangle}{N-1}\big)Fdt$.} We notice that with these strategies, the network topology (small-world or random) is always preserved as time advances.

In the current work, the control parameter of HSP will be the characteristic rewiring frequency parameter $F$, which reflects the time-varying nature of the edges after each (integration) time step $dt$. Larger values of $F$ reflect more rapid switching of the synapses. It is important to note that in real neural networks, the synapses may change at different rates depending on factors such as the developmental stage of the network and/or environmental stimuli. Thus, in the current work, it is reasonable to investigate a large range of rewiring frequencies, $F\in[0.0,1.0]$ \marius{$\mathrm{Hz}$}. However, the rewiring frequencies are typically expected to be small in real neural networks \cite{rakshit2018emergence}. \marius{Thus, the numerical simulations of our HH neural network with small values of $F$ are probably the most relevant indicators of the behavior of CR in real neural networks.}

\R{It is important to understand how STDP affects HSP and vice versa. From the STDP and HSP rules described above, it is clear that the STDP rule, which is controlled by $P$ in our study, does not affect the HSP rule, which is controlled by the rewiring probabilities $(1 - \beta)Fdt$, $\beta Fdt$, and $\big(1-\frac{\langle k\rangle}{N-1}\big)Fdt$. However, HSP can affect  STDP (even at a fixed value of $P$) because the constant swapping of neighbors does not give the newly connected neurons enough time to stabilize their synaptic weight to a saturated value via LTD (since, in our case we have 
$D/P = 1.05>1$ or LTP if we had $D/P<1$). This is why HSP can affect the coupling strength between connected neurons and thus affect synchronization, which can, in turn, affect coherence resonance, as we shall further explain.}

\section{Computational method}\label{Sec. III}
The flow of control in the simulations is presented in the
Appendix. The two outermost loops in the pseudo-code are
on the parameters $P$ and $F$, resulting in Fig. \ref{fig:1}. Each of the other parameters replaces the parameter in the
outermost loop (i.e., $P$) to get results presented in the rest of the figures. 

To measure the degree of regularity of the spiking activity induced by the mechanism of CR in the networks, we use the inverse coefficient of variation --- an important statistical measure based on the time intervals between spikes \cite{pikovsky1997coherence, masoliver2017coherence} and which is related to the timing precision of information processing in neural systems \cite{pei1996noise}. 

For $N=100$ neurons, we numerically integrate Eqs.\eqref{eq:1}-\eqref{eq:5a} with the STDP learning rule of Eq.\eqref{eq:6} and the HSP strategies described above using the Euler–Maruyama algorithm \cite{higham2001algorithmic} with the time step $dt = 0.005$ $\mathrm{ms}$ for a total integration time of $T=2.5\times10^{3}$ $\mathrm{ms}$.  The results shown in the next section were averaged over 20 independent realizations for each set of parameter values and random initial conditions to warrant reliable statistical accuracy with respect to the small-world and random network generations and numerical simulations. For each realization,  we choose random initial points $[V_i(0),x_i(0)]$ for the $i$th ($i = 1,...,N$) neuron with uniform probability in the range of $V_i(0)\in(-75, 40)$, $x_i(0)\in(0,1)$, $x_{i}(0)=\{m_{i}(0),h_{i}(0),n_{i}(0)\}$. As with all the quantities calculated, we have carefully excluded the transient behavior from simulations. After a \marius{sufficiently long transient time of $T_0=2.0\times10^{3}$ $\mathrm{ms}$, we start recording the neuron spiking times $t_i^{\ell}$ ($\ell\in\mathbb{N}$ counts the spiking times)}. 

To prevent unbounded growth, negative conductances (i.e., negative coupling strength), and elimination of synapses (i.e., $g_{ij}=0$), we set a range with the lower and upper  bounds: $g_{ij}\in[g_{min},g_{max}]=[0.0001,0.35]$, \marius{where $g_{max}=0.35$ $\mathrm{mS/cm^2}$ is in the range of the maximum synaptic conductances 
[0.3,0.5] $\mathrm{mS/cm^2}$ usually measured in the standard Hodgkin-Huxley neuron \cite{ren2012hopf,popovych2013self}, and the lower bound $g_{min}=0.0001$ is set to ensure not to miss any effects that occur outside of classical parameter ranges.} Moreover, the initial weight of all excitable synapses is normally distributed in the interval $[g_{min},g_{max}]$, with mean $g_0=0.185$ and standard deviation $\sigma_0=0.02$.

For $N=100$ coupled neurons, \marius{the reciprocal  of the normalized standard deviation of the mean inter-spike intervals, denoted here by $\Omega$, measures the average variability of spike trains of the network \cite{gabbiani1998principles} (i.e., a measure of the regularity of the noise-induced spiking activity), and is computed as \cite{masoliver2017coherence,gong2005optimal}:}
\begin{equation}\label{eq:9}
    \Omega =
    \frac{\overline{\langle \mathrm{\tau} \rangle}}{\sqrt{\overline{\langle \mathrm{\tau}^2 \rangle} - \overline{\langle \mathrm{\tau} \rangle^2}}},
\end{equation}
where
$\overline{\langle \mathrm{\tau} \rangle} = N^{-1} \sum_{i=1}^N \langle \mathrm{\tau}_i \rangle$ and 
$\overline{\langle \mathrm{\tau}^2 \rangle} = N^{-1}\sum_{i=1}^N \langle \mathrm{\tau}_i^2 \rangle$, with
$\langle \mathrm{\tau}_i \rangle$ 
and 
$\langle \mathrm{\tau}_i^2 \rangle$ representing
the mean and mean squared \marius{inter-spike interval} (over time),  $\mathrm{\tau}_i = t_i^{\ell+1}-t_i^{\ell}>0$, of neuron $i$.
We determine the spike occurrence times $t_i^{\ell}$ from the instant $t$ when the membrane potential variable $V_i(t)$ crosses the threshold $V_{\mathrm{th}}=0.0$ \marius{$\mathrm{mV}$}. A larger (lower) value of $\Omega$ indicates a higher (lower) degree of CR, i.e., a higher (lower) average temporal coherence \marius{of the spiking neurons.}
\marius{At this point, we emphasize the fact that $\Omega$ is just a measure of the regularity of the noise-induced spiking activity and not a measure of the efficiency of the control method, which compares the relative time scale of the neuronal dynamics and the networks updates to the total number of time steps in the integration. The occurrence of coherent noise-induced spiking activity, i.e., CR, crucially depends on whether the system's parameters are fixed sufficiently near but before a bifurcation threshold. How coherent or incoherent (as measured by $\Omega$) the noise-induced spiking activity is,  depends on how close the parameters are to the bifurcation threshold. Varying one or some of these parameters (e.g., $\tau_c$, $F$) will vary the system's proximity from the bifurcation thresholds and, consequently, a variation in the degree of coherence indicated by the variation in $\Omega$.}

\marius{Furthermore, we notice that the network size considered in this work ($N=100$) is significantly smaller than those of the real neural networks in the brain --- the order of 80 billion neurons. Thus, the results presented in this work are a test of principle. For the moment, we will only focus on whether and, if so, how STDP and HSP can jointly affect the degree of CR. Future research projects could investigate large-network-size effects on the degree of CR in the presence of both STDP and HSP.}

\begin{figure*}
\centering
\includegraphics[width=8.0cm,height=4.0cm]{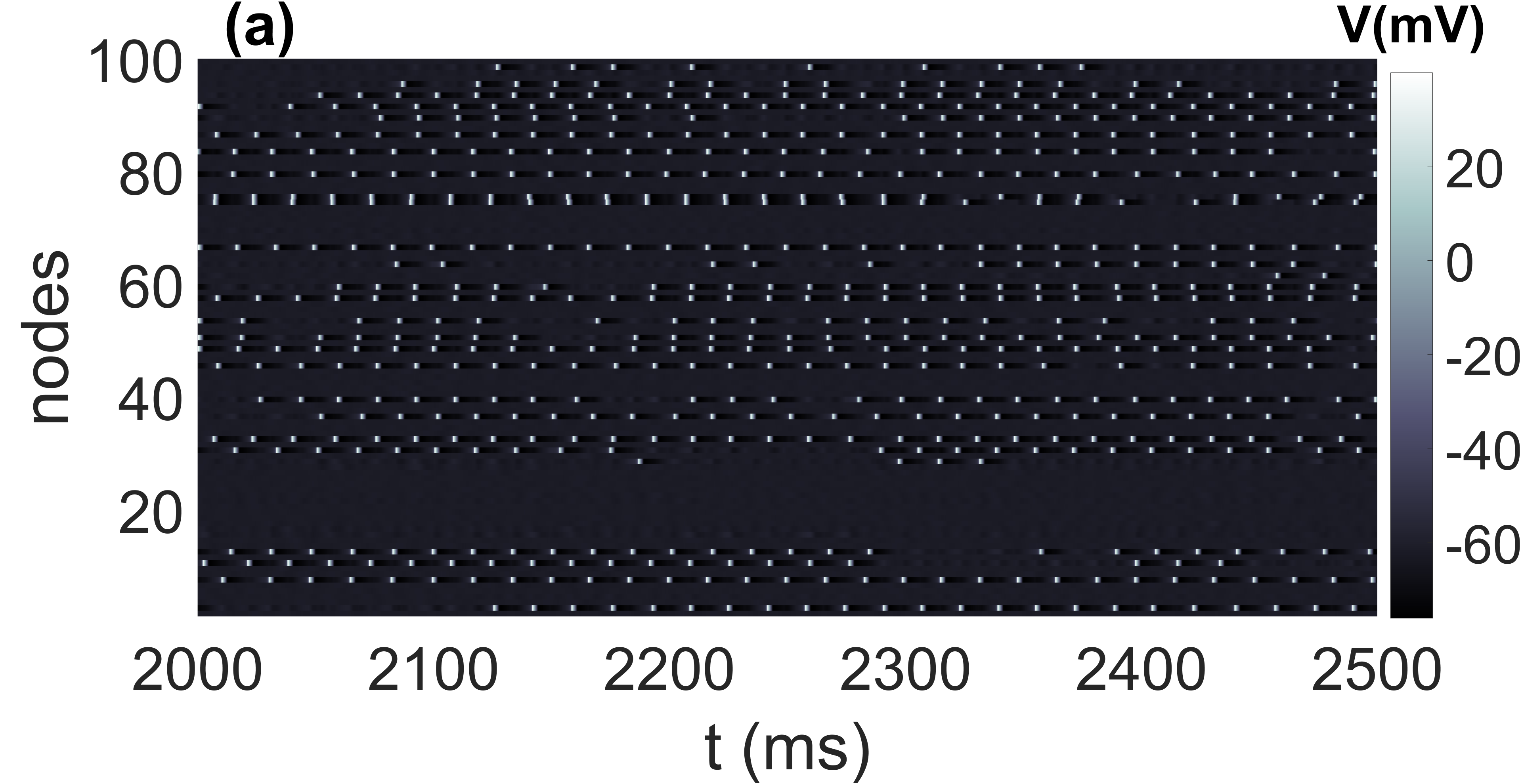}\includegraphics[width=8.0cm,height=4.0cm]{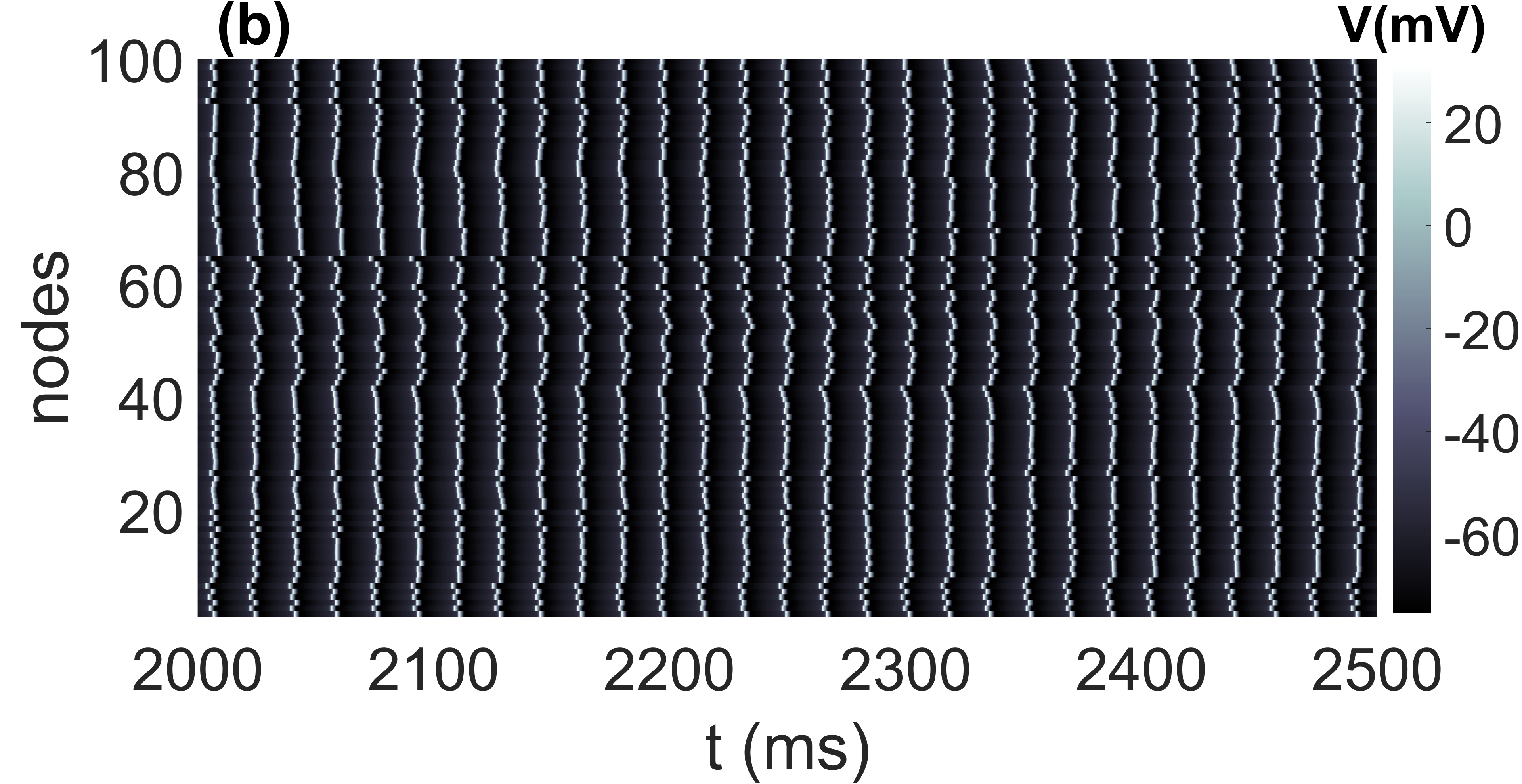}\\[3.0mm]\includegraphics[width=8.0cm,height=4.0cm]{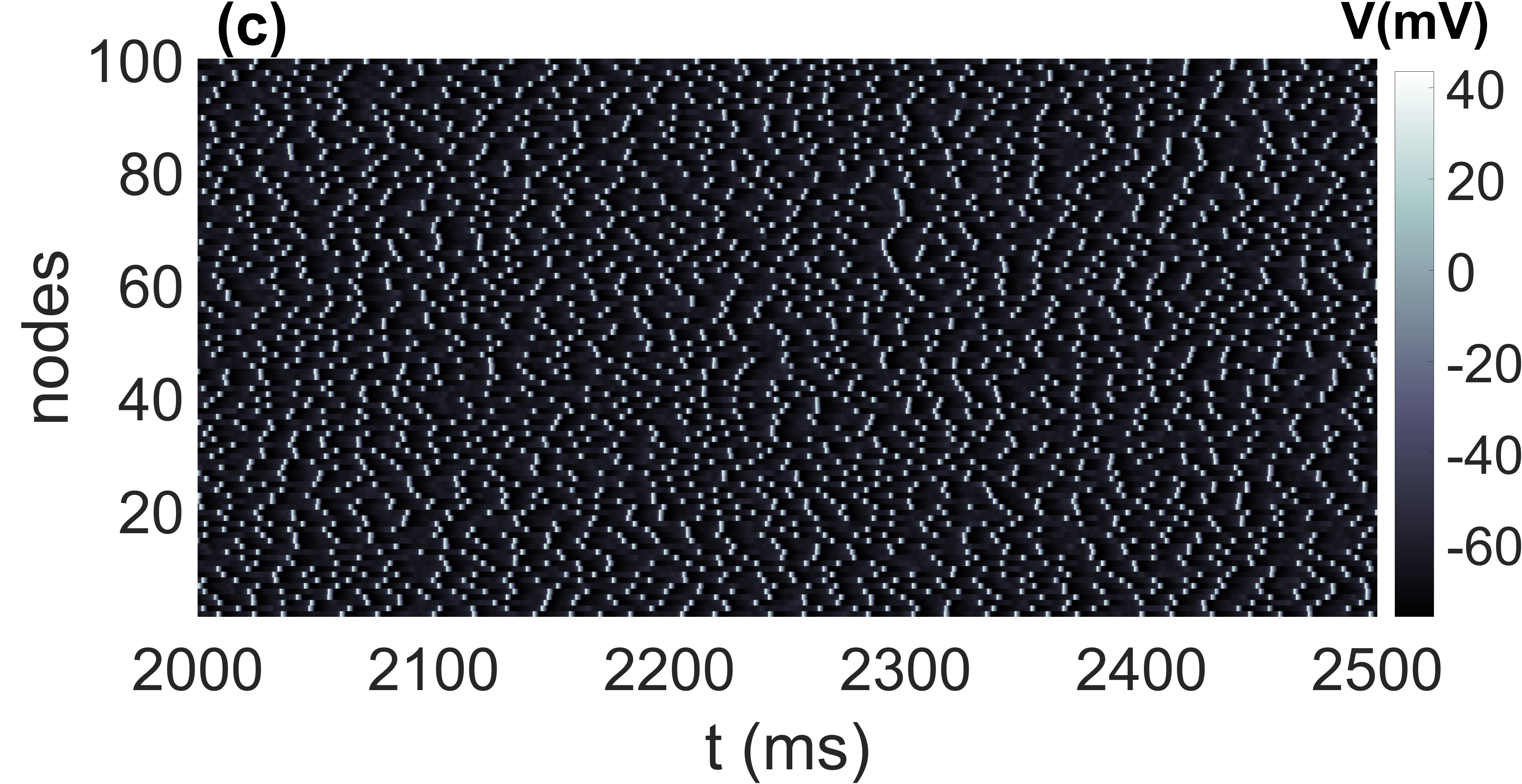}\includegraphics[width=8.0cm,height=4.0cm]{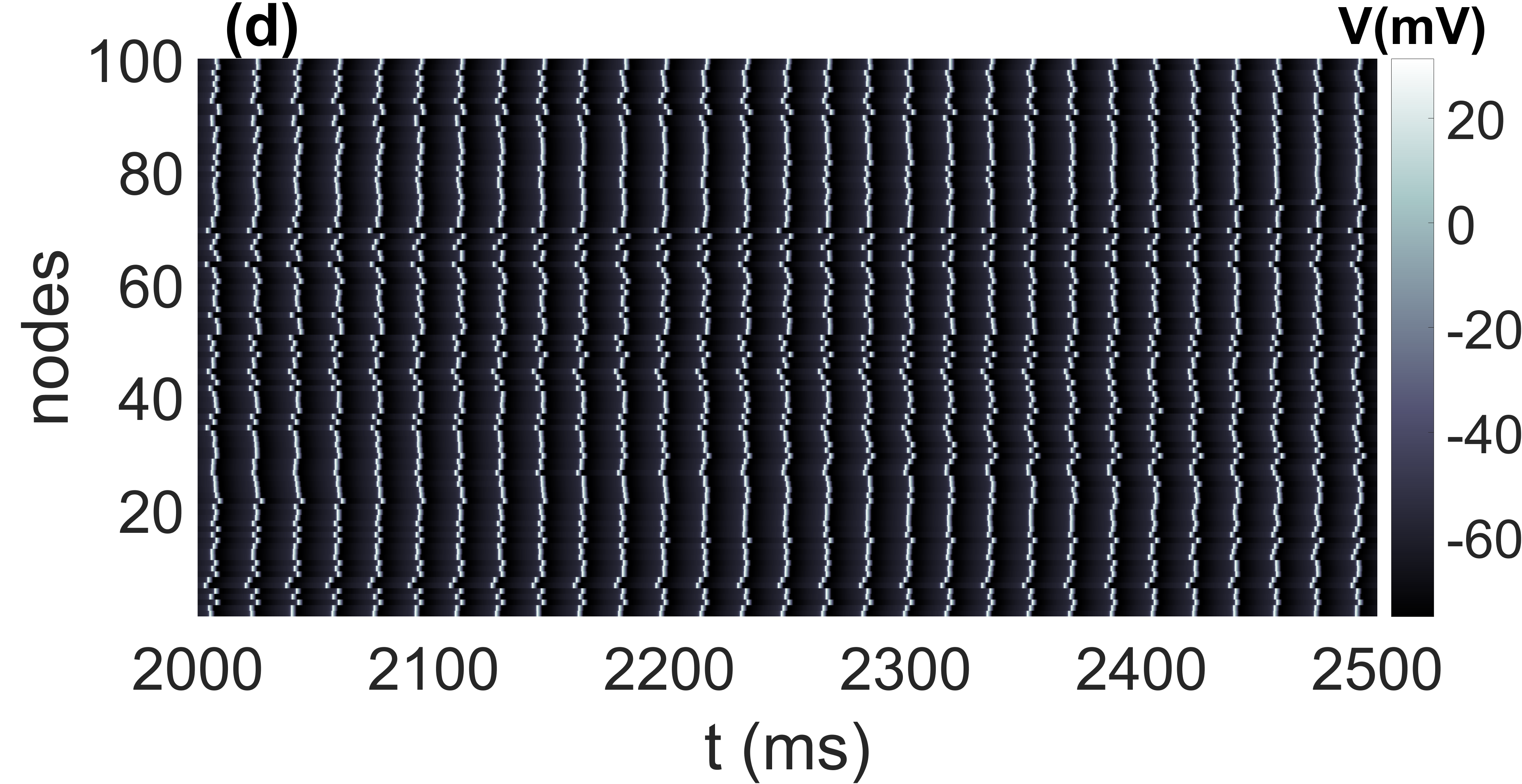}
\caption{\marius{Spatio-temporal activity of the membrane potential variable $V$ (in $\mathrm{mV}$) in (a)-(c) small-world network with $\beta=0.25$, and (d) random network with $\beta=1$. (a) $A=400 $ $\mathrm{\mu m^2}$: incoherent activity \R{with $\Omega=7.56$}. (b) $A=4.0 $ $\mathrm{\mu m^2}$: coherent activity \R{with $\Omega=54.10$} and  period of oscillation $\overline{\tau}=15.95$ $\mathrm{ms}$. (c) $A=0.15$ $\mathrm{\mu m^2}$: incoherent activity \R{with $\Omega= 2.12$.} (d) $A=4.0 $ $\mathrm{\mu m^2}$: coherent activity \R{with $\Omega= 54.56$} and period of oscillation $\overline{\tau}=15.95$ $\mathrm{ms}$.
Other parameters: $P=0.1\times10^{-5}$, $F=1.0\times10^{-3}$ $\mathrm{Hz}$, $\tau_c=13.0$ $\mathrm{ms}$, $\langle k \rangle=5$.}}
\label{fig:0}
\end{figure*}

\marius{In Figs.~\ref{fig:0}(a)-(c), we show examples of the neural activity in a small-world ($\beta=0.25$) neural network consisting of $N=100$ neurons while also illustrating the phenomenon of CR with respect to the channel noise intensity, which is controlled by the membrane patch area $A$. We notice from Eq. \eqref{eq:4} that membrane patch area $A$ appears in the denominator. Thus, the intensity of channel noise is inversely proportional to the membrane patch area $A$ --- the larger (smaller) $A$ is, the weaker (stronger) the channel noise intensity. With smaller $A$, ions scramble for the few available channels. In comparison, with larger $A$, their movements become more deterministic as a sufficiently large number of opened channels are now available.} 

\marius{In  Fig. \ref{fig:0}(a), we apply a weak channel noise intensity by fixing the membrane patch area at a large number, $A=400 $ $\mathrm{\mu m^2}$. The neural activity indicates incoherent spiking \R{(with a low inverse coefficient of variation of $\Omega=7.56$)}, with some neurons even incapable of spiking. In  Fig. \ref{fig:0}(b), we increase the membrane patch area to $A=4.0$ $\mathrm{\mu m^2}$ and the neural activity achieves resonance during which the spiking becomes very coherent \R{(with a high $\Omega= 54.10$)} and a period (average inter-spike interval) of $\overline{\tau}=15.95$ $\mathrm{ms}$. In  Fig. \ref{fig:0}(c), when we apply a strong channel noise by fixing $A=0.15$ $\mathrm{\mu m^2}$, the neural activity becomes very incoherent \R{(with a very low $\Omega= 2.12$)}. In  Fig. \ref{fig:0}(d), we show an example of the neural activity in the random ($\beta=1$) neural network at peak coherence  \R{(with a high $\Omega= 54.56$)} and a period (average inter-spike interval) of $\overline{\tau}=15.95$ $\mathrm{ms}$.}

\marius{As we pointed out earlier, CR occurs when the regularity of noise-induced oscillations is a non-monotonic function of the noise amplitude (in our case $A$) or another system parameter. So, throughout the rest of this paper, we fix the membrane patch area at its resonant value, i.e., $A=4.0.$ $\mathrm{\mu m^2}$, and study CR with respect to  the STDP, HSP, and the network parameters ($P$, $F$ $\tau_c$, $\langle k \rangle$, $\beta$).}

\marius{Furthermore, before presenting and discussing the results, it is important to point out already that the spiking frequency at resonance in the networks in Fig. \ref{fig:0} is given by \R{$1/\overline{\tau}\approx63$} $\mathrm{Hz}$ can be lost when the system parameters change and  push the system into a stronger excitable regime  leading to incoherent oscillations like in Figs. \ref{fig:0}(a) and (c). Therefore, unlike what one would most probably expect,  the observed behaviors of the degree of CR do not emerge due to the interplay between the rewiring frequency and the spiking frequency at resonance (which may even be lost for certain parameter values). Hence, we provide theoretical explanations for our results from the perspective of phase synchronization.}

\section{Numerical results and discussion}\label{Sec. IV}
We recall that we aim to study the combined effect of the HSP strategy (controlled by the characteristic rewiring frequency parameter $F$) and (i) the STDP rule (controlled by the adjusting rate parameter $P$), (ii) the time delay of chemical synapses $\tau_c$, (iii) the average degree of the networks $\langle k \rangle$, (iv) the rewiring probability of the network $\beta$, on the degree of coherence of spiking in small-world and random networks. 

\subsection{Combined effects of $F$ and $P$}
\begin{figure*}
\centering
\includegraphics[width=5.5cm,height=3.41cm]{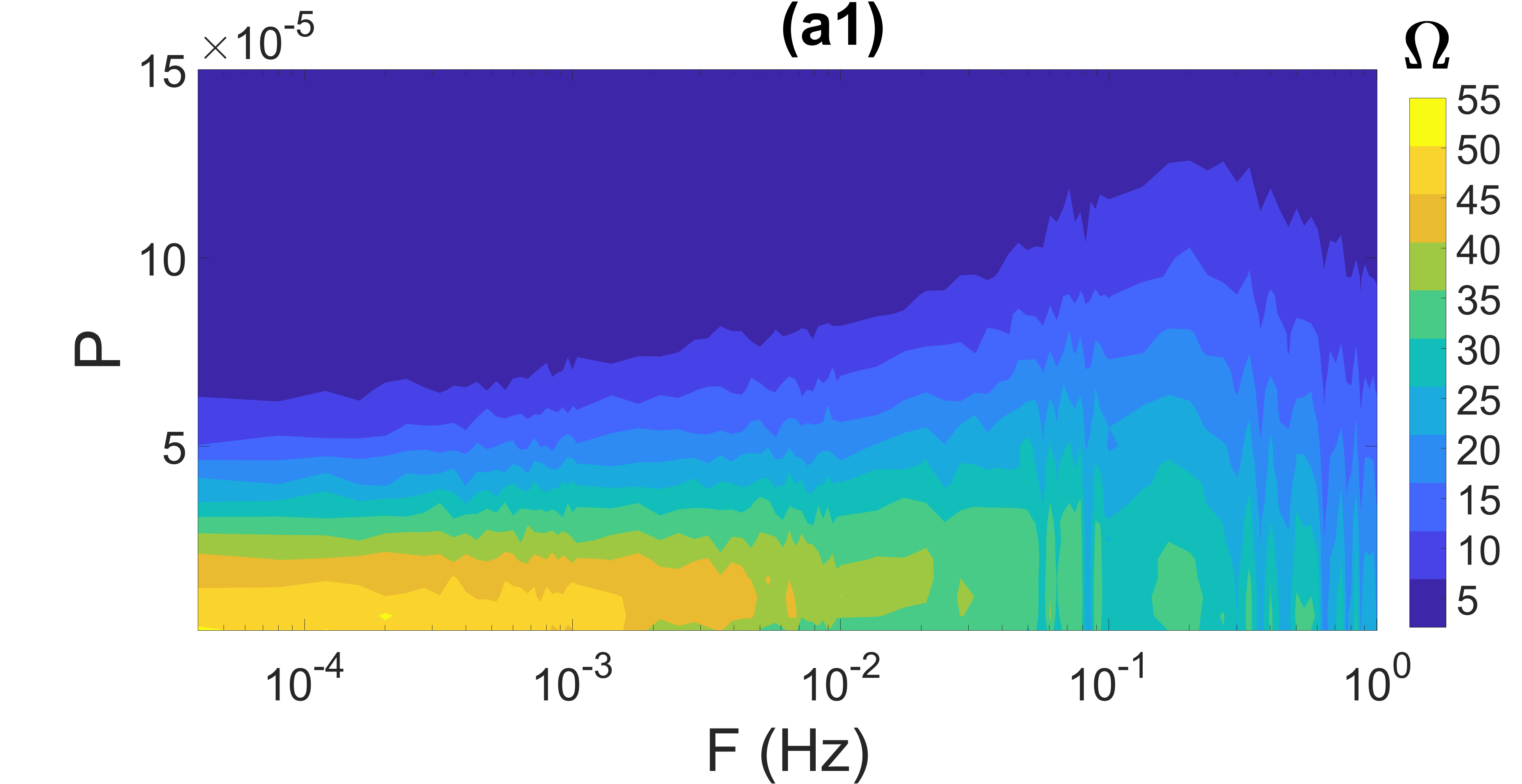}\includegraphics[width=5.5cm,height=3.41cm]{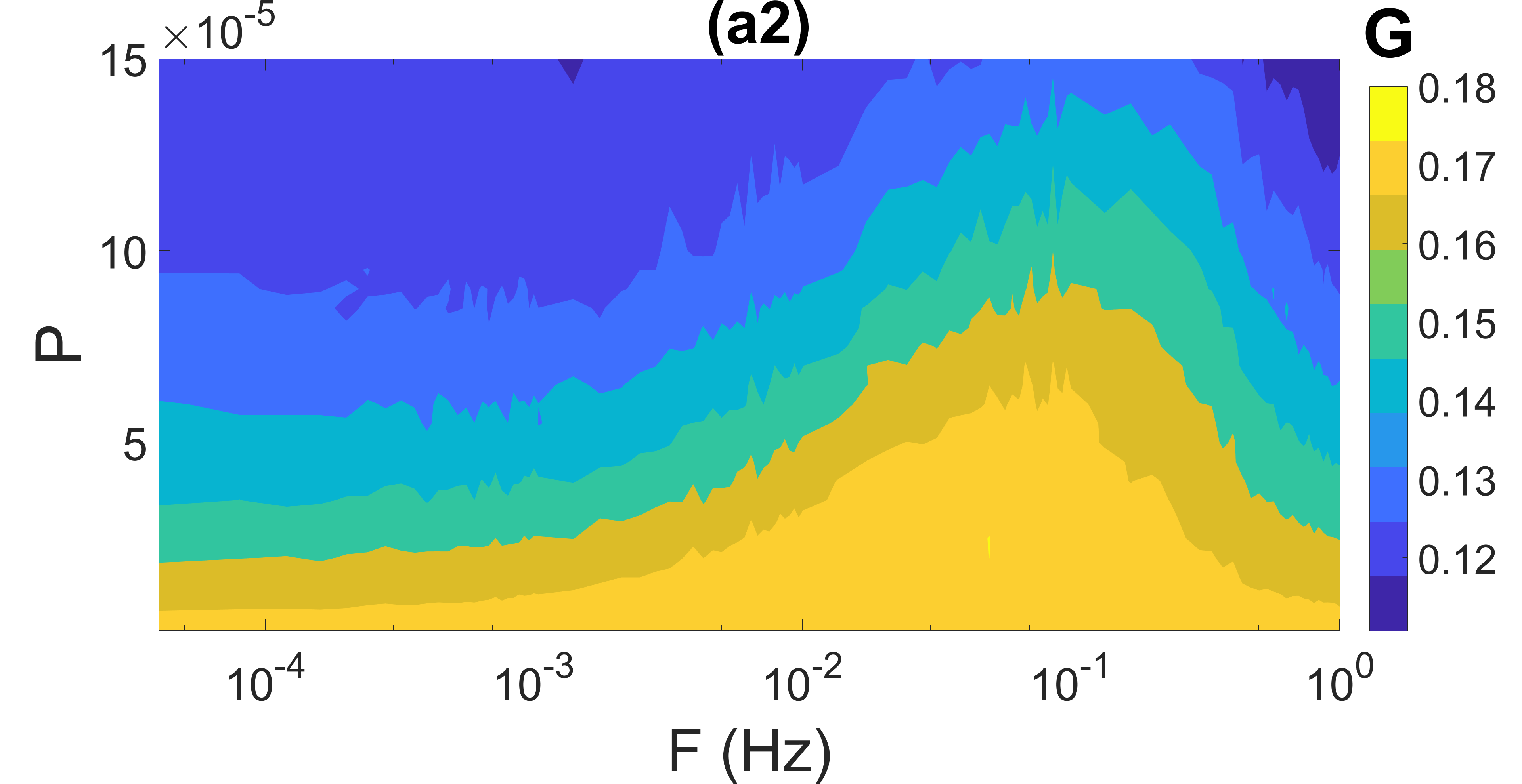}\includegraphics[width=5.5cm,height=3.41cm]{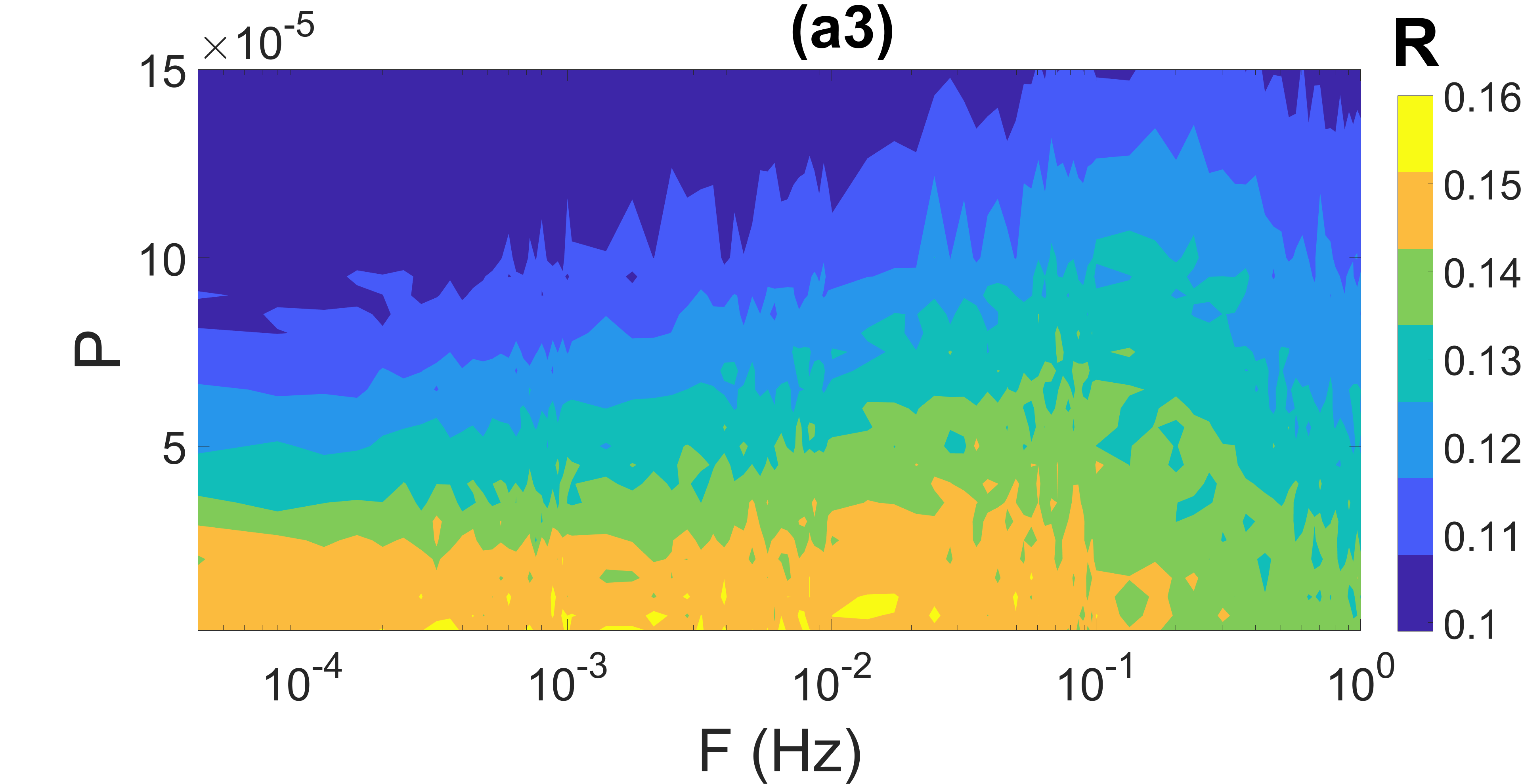}\\[2.0mm]
\includegraphics[width=5.5cm,height=3.41cm]{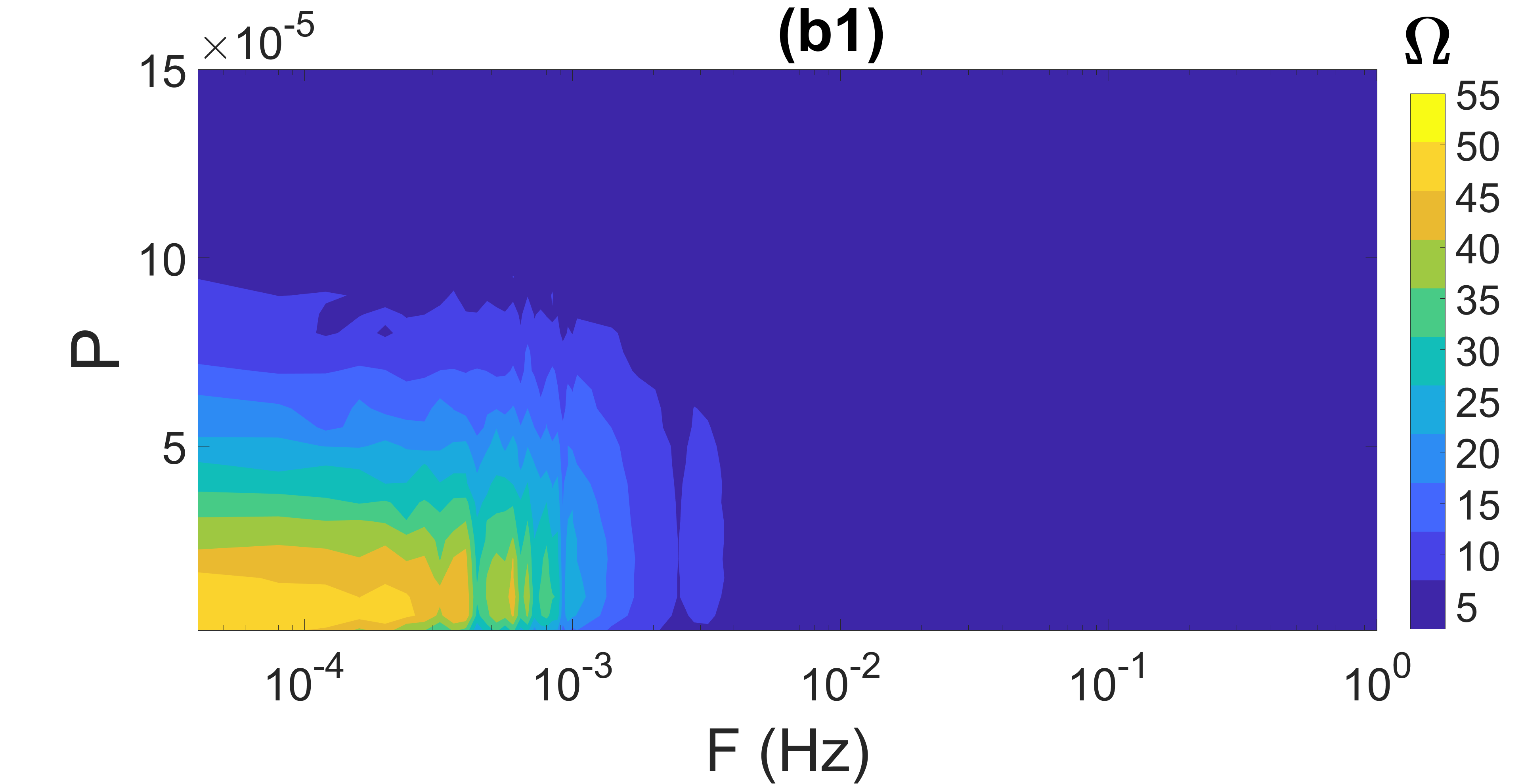}\includegraphics[width=5.5cm,height=3.41cm]{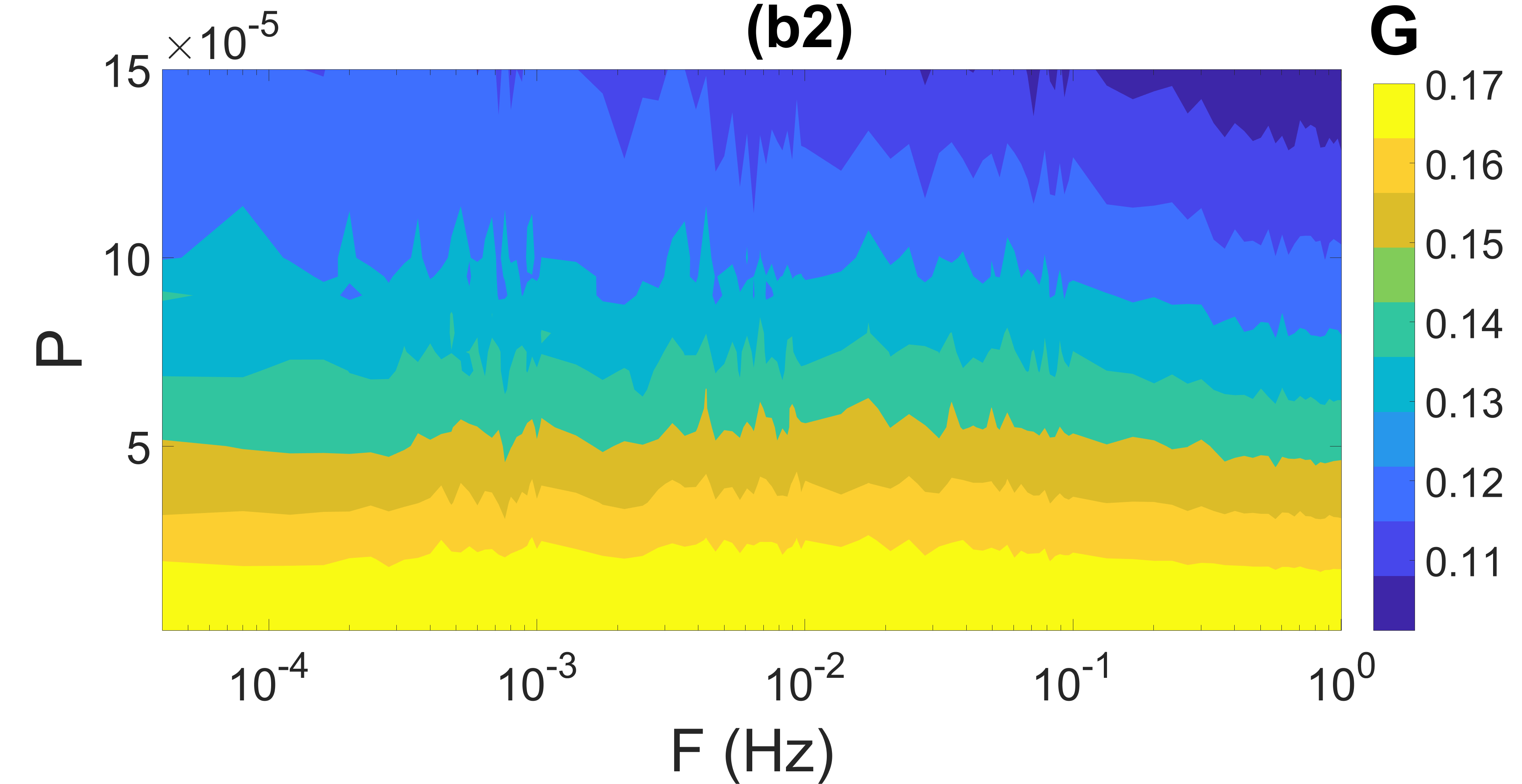}\includegraphics[width=5.5cm,height=3.41cm]{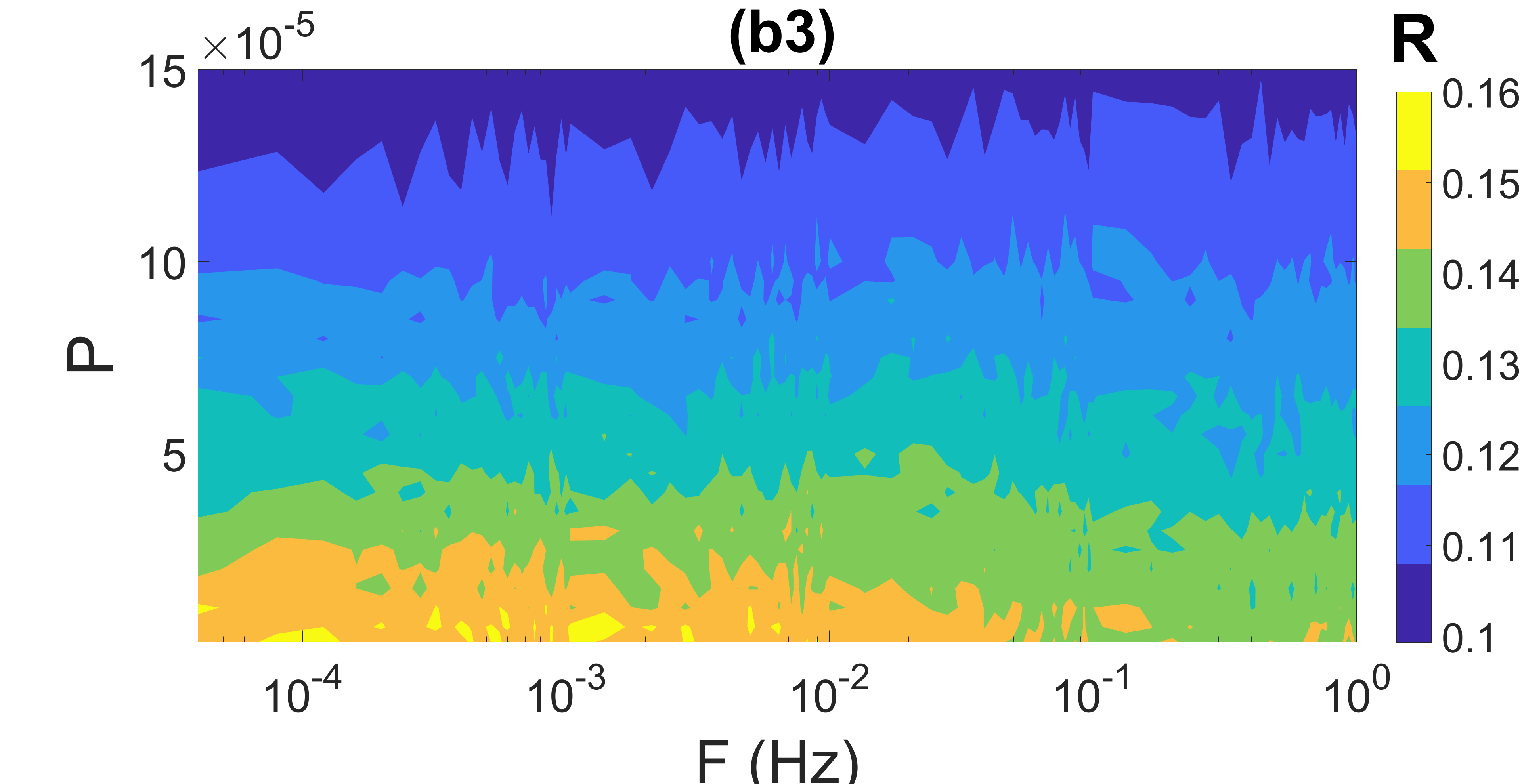}
\caption{Inverse coefficient of variation $\Omega$ in the rewiring frequency $F$ and adjusting rate $P$ parameter space for: (a1) small-world network with $\beta=0.25$, $\langle k\rangle=5$, $\tau_c=13.0$ \marius{$\mathrm{ms}$}, and (b1) random network with $\beta=1$, $\langle k\rangle=5$, $\tau_c=13.0$ \marius{$\mathrm{ms}$}. 
The corresponding variations of the average coupling strength $G$ in the networks in (a1) and (b1) are shown in (a2) and (b2), respectively. The corresponding variation in the degree of phase synchronization $R$ in the networks in (a1) and (b1) are shown in (a3) and (b3), respectively.}
\label{fig:1}
\end{figure*}

In Figs. \ref{fig:1}(a1)-(a3), we respectively depict in the $F-P$ plane the contour plots of $\Omega$ measuring the degree of CR, the average coupling strength of the network $G$, and the degree of phase synchronization $R$ for a time-varying small-world network with time delay $\tau_c=13.0$ \marius{$\mathrm{ms}$}, average degree $\langle k \rangle=5$, and rewiring probability $\beta=0.25$. To see how the average coupling strength $G\in[0.0001,0.35]$ and the degree of phase synchronization $R\in[0,1]$ of the network changes with $F$ and $P$, we will average the synaptic weights and the Kuramoto order parameter over the entire population and time:
\begin{equation}\label{eq:8}
\small{
G = \displaystyle{\Bigg \langle\frac{1}{N^2}\sum\limits_{i=1}^{N}\sum\limits_{j=1}^{N}g_{ij}(t)\Bigg \rangle_t};\:
R=\displaystyle{\Bigg \langle\bigg|\frac{1}{N}\sum\limits_{j=1}^N\exp{[i\phi_{j}(t)]}\bigg|\Bigg \rangle_t}}
\end{equation}
 where $\big|\cdot\big|$ represents the absolute value and $\big\langle\cdot\big\rangle_t$ the average over time in the interval $t\in[T_0,T]$. In the argument of the exponential function: $i = \sqrt{-1}$ and the quantity $\phi_{j}(t)= 2\pi \ell + 2\pi(t-t_{_{j}}^{\ell})/(t_{_{j}}^{\ell+1}-t_{_{j}}^{\ell})$ with $t\in[t_{_{j}}^{\ell},t_{_{j}}^{\ell+1})$ approximates the phase of the $j$th neuron and linearly increases over $2\pi$ from one spike to the next. The degree of phase synchronization increases as $R\in[0,1]$ increases.

As we pointed out earlier, when the effect of depression dominates that of potentiation (i.e., when $D$ is larger than $P$), the decrement of synaptic strength is stronger than its increment. Hence, the smaller the value of $P$, the larger the network's average coupling strength $G$.

In Fig. \ref{fig:1}(a1), the contour of $\Omega$ is plotted against $F$ and $P$ for a small-world network. The result indicates that for static (i.e., when the rewiring frequency is $F=0$ \marius{$\mathrm{Hz}$} and slowly varying small-world networks (i.e., when $F\in[0,0.5\times10^{-3}]$ \marius{$\mathrm{Hz}$}), $\Omega$ increases (indicating a higher degree of coherence) as $P$ decreases. For smaller values of $P (<1.0\times10^{-5})$, as the rewiring frequency increases (i.e., when $F>0.5\times10^{-3}$ \marius{$\mathrm{Hz}$}),  $\Omega$ decreases (indicating lower degree of coherence).

In Fig. \ref{fig:1}(a2), the contour of $G$ is plotted against $F$ and $P$ for the small-world network of Fig. \ref{fig:1}(a1). 
First, because $P$ is smaller than $D$ (i.e., $D/P=1.05$), the depression dominates the synaptic modifications as the 
average coupling strength $G\in[0.11,0.18]$ is always less than the network's mean initial synaptic strength $g_0=0.185$. We observe that at each value of $F$, $G$ always increases as $P$ decreases. And for each value of $P$, as $F$ increases in $[0,1]$, $G$ shows a non-monotonic behavior with a maximum value occurring at $F\approx10^{-1}$ \marius{$\mathrm{Hz}$}. 

In Fig. \ref{fig:1}(a3), the contour of $R$ is plotted against $F$ and $P$ for the small-world network of Fig. \ref{fig:1}(a1) and whose average coupling strength is depicted in Fig. \ref{fig:1}(a2). We observe that the degree of phase synchronization $R$ increases with decreasing $P$ --- an immediate consequence of the average coupling strength $G$ increasing with decreasing $P$. We observe that at each value of $F$, $R$ increases as $P$ decreases. And for each value of $P$, as $F$ increases in $[0,1]$, $R$ also shows a non-monotonic behavior with a maximum value occurring at $F\approx10^{-1}$ \marius{$\mathrm{Hz}$}, just as with $G$ in Fig. \ref{fig:1}(a2). It is worth noting that the highest degree of phase synchronization achieved is never full, i.e., the largest value of the order parameter is $R\approx0.16\neq1$. 
\marius{This is because LTD dominates (as a result of setting $D/P=1.05$) in the network, with the average synaptic weight between the neurons weakening below $g_{max}/2$.} Nevertheless, the network's synchronization degree is strong enough to affect the degree of CR. 

In Figs. \ref{fig:1}(b1)-(b3), we present the contours of $\Omega$, $G$, and $R$ for the random network, respectively. First, we notice that similar to a small-world network, smaller $P$ (and consequently larger $G$ and $R$) and smaller $F(<1.5\times10^{-4}$ \marius{$\mathrm{Hz}$}), the random network produces a larger $\Omega$ (i.e., higher coherence). The differences in the behavior of coherence in both types of networks are that: (i) In the random network, independently of the value of $P$ (and consequently, independently of the value of $G$), higher rewiring frequencies ($F>10^{-3}$ \marius{$\mathrm{Hz}$}) permanently deteriorate the degree of coherence, as indicated by the very low value of $\Omega\approx0.10$. On the other hand, in the small-world network, we can still have a high degree of coherence with $\Omega\ge30$, depending on the value of $P<2.5\times10^{-5}$. (ii) In the random network, a significantly slower switch of synaptic connections, i.e., when $F<1.5\times10^{-4}$ \marius{$\mathrm{Hz}$}  (compared to $F<0.5\times10^{-3}$ \marius{$\mathrm{Hz}$} in the small-world network) is optimal for the best degree of coherence occurring in the \R{light yellow regions} of Figs. \ref{fig:1}(a1) and (b1).

Comparing Figs. \ref{fig:1}(a1)-(a3) with a small-world topology, we observe that in general, $\Omega$ increases to its highest values as $G$ increases (causing $R$ to also increase), especially at smaller values of $F<0.5\times10^{-3}$ \marius{$\mathrm{Hz}$}. For a random network in Figs. \ref{fig:1}(b1)-(b3), with even smaller values of $F<1.5\times10^{-4}$ \marius{$\mathrm{Hz}$}, a similar behavior is observed. 

The fact that the degree of coherence becomes better at larger $G$ (i.e., smaller $P$) and smaller $F$ in both small-world and random networks can be intuitively explained as follows: First, we recall that the chemical synapses are uni-directional and so information from the pre-synaptic neuron $j$ can be transferred to the post-synaptic neuron $i$, \marius{but not the other way around} (as it will have been if bi-directional electrical synapses mediated the links). Secondly, the noise in the network is local, i.e., $\xi_{x_i}(t)$ ($i=1,2,...,N$) in Eq.\eqref{eq:2} are independent Gaussian processes. The locality of these stochastic processes naturally introduces some heterogeneity in the noise-induced spiking times of the neurons in the network. Thirdly, we also note that connected neurons stay permanently connected (i.e., when $F=0$ \marius{$\mathrm{Hz}$}) or remain connected for a relatively long time (i.e., when $0<F\ll1$ \marius{$\mathrm{Hz}$}) before switching their connections to previously unconnected neurons. 

 In both cases, the better degree of synchronization \R{[light and dark yellow regions in Fig. \ref{fig:1}(a3) bounded by $F\in[0,10^{-1}]$ \marius{$\mathrm{Hz}$} and $P<2.5\times10^{-5}$ and in Fig. \ref{fig:1}(b3) bounded by $F\in[0,10^{-1})$ \marius{$\mathrm{Hz}$} and $P<1.5\times10^{-5}$ ]} induced by larger values of $G$ \R{[dark yellow regions in Fig. \ref{fig:1}(a2) bounded by, e.g., $P<1.0\times10^{-5}$ when $F\in[0,10^{-3}]$ and 
 $P<5.0\times10^{-5}$ when $F\in[10^{-2},10^{-1}] $\marius{$\mathrm{Hz}$} and the light yellow in Fig. \ref{fig:1}(b2) bounded by $F\in[0,10^{0}]$ \marius{$\mathrm{Hz}$}  and $P<1.6\times10^{-5}$]} is maintained for a relatively long time ($0<F\ll1$ \marius{$\mathrm{Hz}$}) or permanently ($F=0$ \marius{$\mathrm{Hz}$}).

Now, this relatively stronger degree of synchronization may occur via two scenarios: (I) the post-synaptic neurons $i$ with \textit{less coherent} spiking times synchronizing the pre-synaptic neurons $j$ with \textit{more coherent} spiking times. This would then lead to an overall  \textit{more coherent} spiking times of the entire network as indicated by the \R{light yellow regions} in Figs. \ref{fig:1}(a1) and (b1) with a higher degree of coherence $\Omega\ge50$. (II) the post-synaptic neurons $i$ with \textit{more coherent} spiking times synchronizing the pre-synaptic neurons $j$ with \textit{less coherent} spiking times. In this case, we get an overall  \textit{less coherent} spiking times of the entire network as indicated by the lower degree of coherence with \R{$\Omega<40$ represented by all the non-yellow colors in Fig. \ref{fig:1}(a1) (i.e., when $F>0.5\times10^{-3}$ \marius{$\mathrm{Hz}$}) and Fig. \ref{fig:1}(b1) (i.e., when $F>1.5\times10^{-4}$ \marius{$\mathrm{Hz}$}).}

In Figs. \ref{fig:1}(a1) and (b1), we observe a deterioration of the degree of coherence (especially in the random network in Figs. \ref{fig:1}(b1)) for small $P<2.5\times10^{-5}$ and higher frequencies ($F>10^{-3}$ \marius{$\mathrm{Hz}$}), \marius{even though for} this same range of values of $P$ and $F$, the average coupling strength $G$ is relatively strong (see Figs. \ref{fig:1}(a2) and (b2)) leading to the relatively high degree synchronization (see Figs. \ref{fig:1}(a3) and (b3)) observed when the networks were static or slowly varying. \marius{This observation can be explained by the occurrence of synchronization via scenario II described above, in addition to the fact that this degree of synchronization is unstable due to rapidly switching links between the neurons.}

\marius{In Figs. \ref{fig:1}(a3) and (b3), we can observe some fluctuations, i.e., several small \R{light yellow regions in the dark yellow region}. To explain this observation, it is worth noting that the numerical difference between the values of $R$ in the \R{light yellow and dark yellow areas} is minimal. Secondly, because the synaptic weights exhibit long-term depression (i.e., weakening of synaptic strength as $P$ decreases), it becomes harder for now weakly coupled neurons to (phase) synchronize their noise-induced spiking activity. Hence, the average over the number of realizations of the simulations sometimes leads to minimally different values of the order parameter $R$ as $P$ and $F$ change.}

In the remaining subsections of this paper, we study the effects of HSP and STDP on CR by looking at the behavior of $\Omega$ with respect to $F$ at three different values of $P$ when the network control parameters, including the time delay of the connections $\tau_c$, the average degree $\langle k \rangle$, and the rewiring probability $\beta$ of the networks, are independently varied. The three values of $P$ selected are such that we have (i) a large average synaptic strength G, i.e., $P=0.1\times10^{-5}$, leading to a relatively high degree of synchronization, (ii) an intermediate average synaptic strength G, i.e., $P=3.0\times10^{-5}$, leading to an intermediate degree of synchronization, and (iii) a small average synaptic strength G, i.e., $P=12.5\times10^{-5}$, leading to a low degree of synchronization.

\subsection{Combined effect of $F$, $\tau_c$, and $P$}
\begin{figure*}
\centering
\includegraphics[width=5.5cm,height=3.41cm]{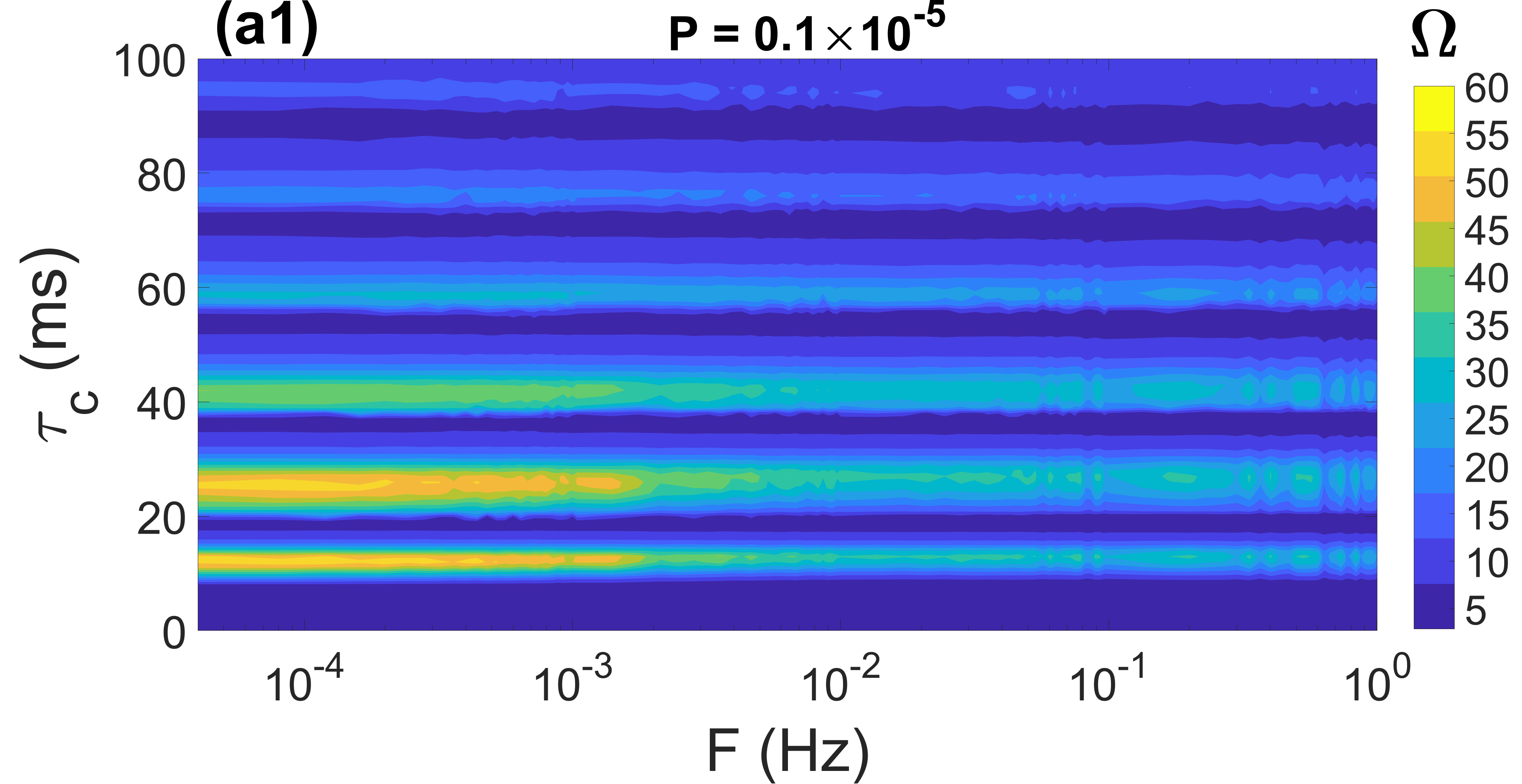}\includegraphics[width=5.5cm,height=3.41cm]{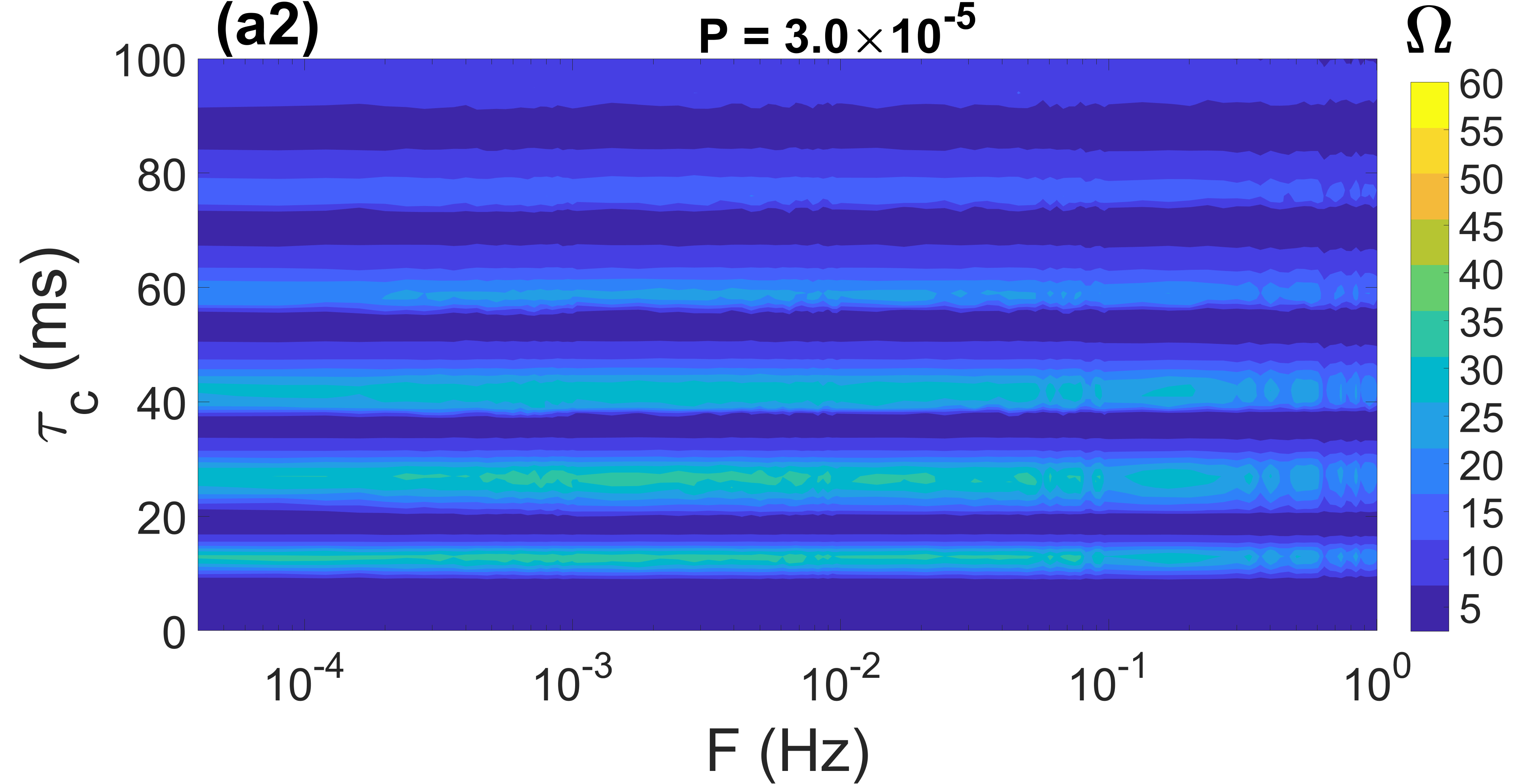}\includegraphics[width=5.5cm,height=3.41cm]{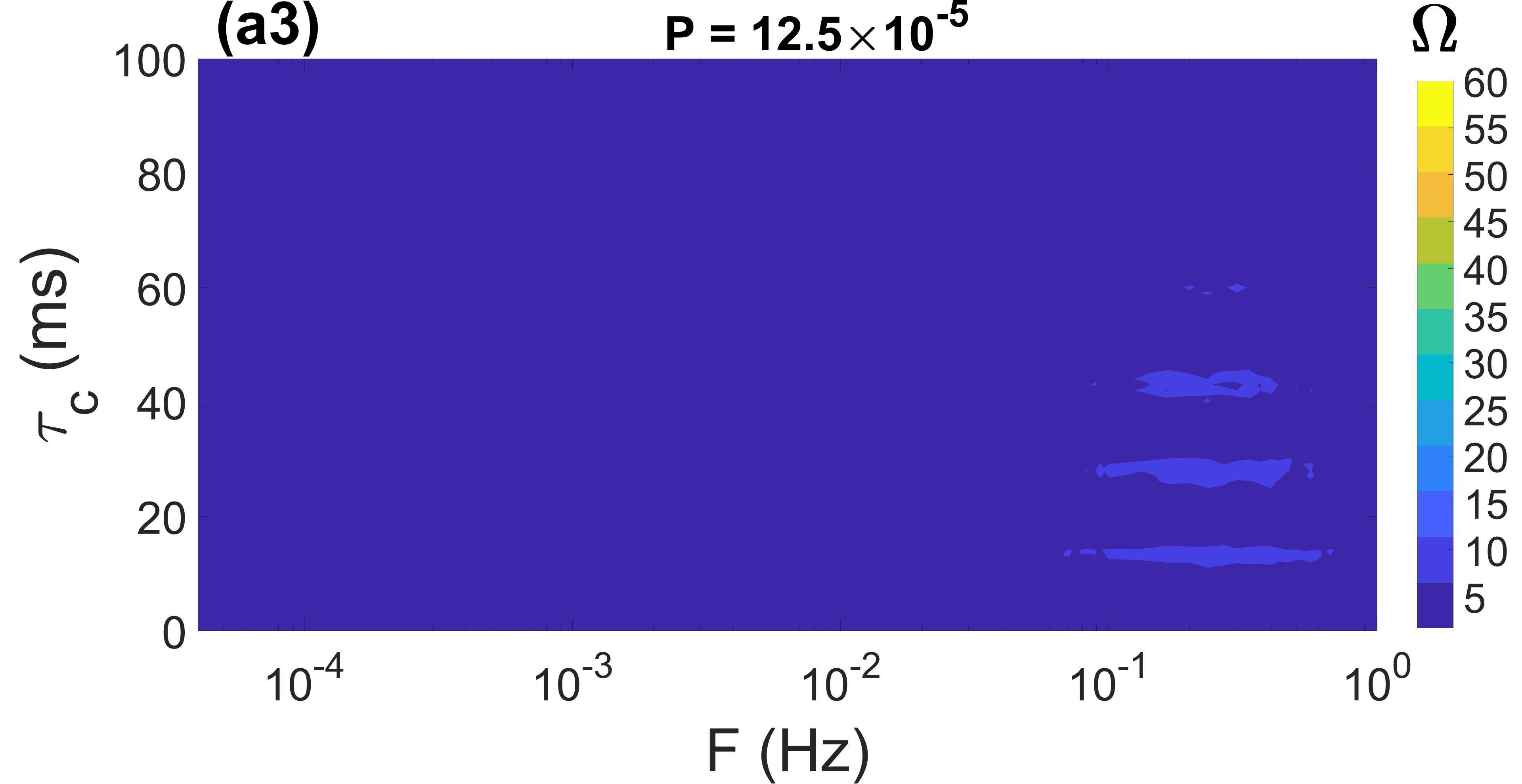}\\[2.0mm]
\includegraphics[width=5.5cm,height=3.41cm]{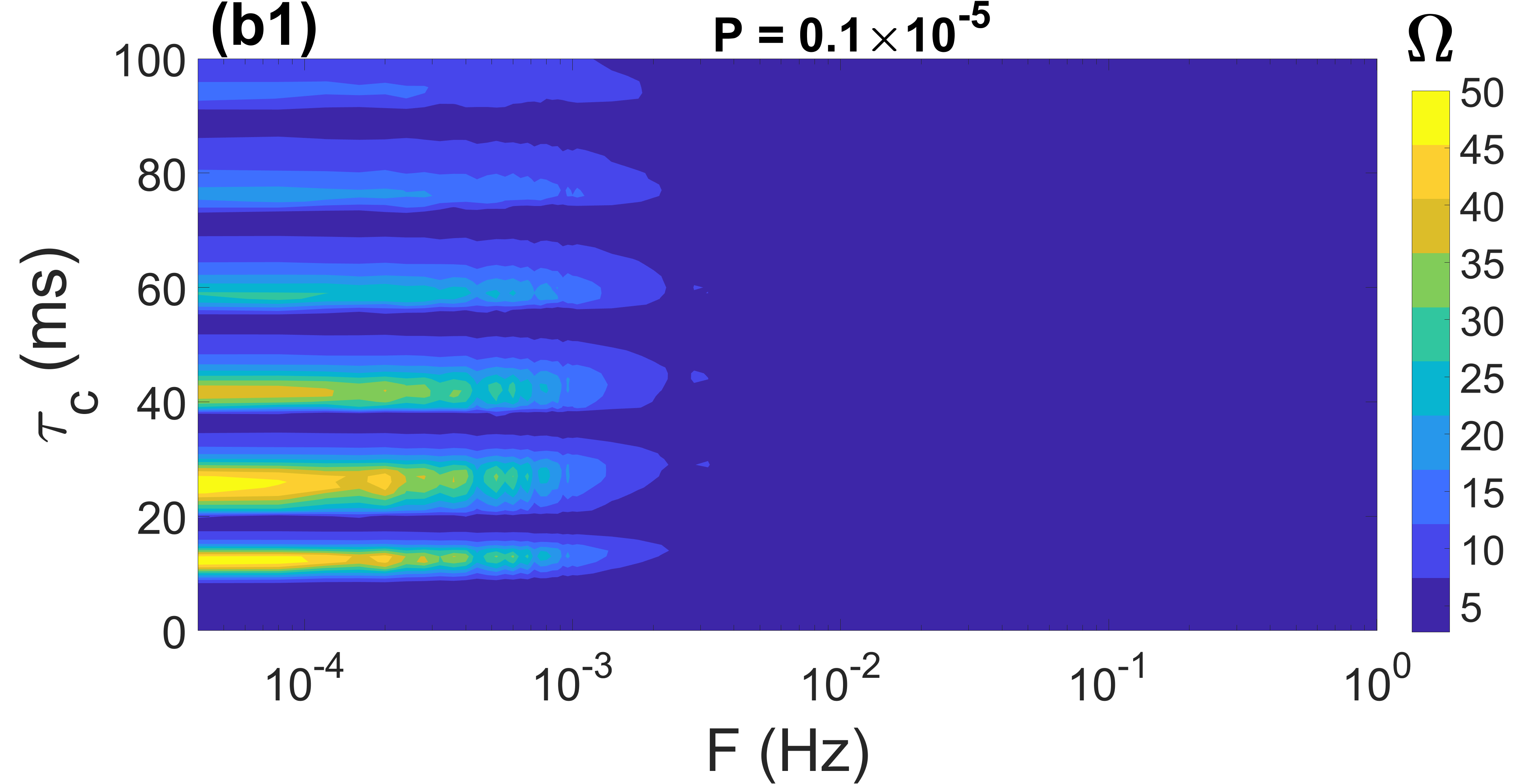}\includegraphics[width=5.5cm,height=3.41cm]{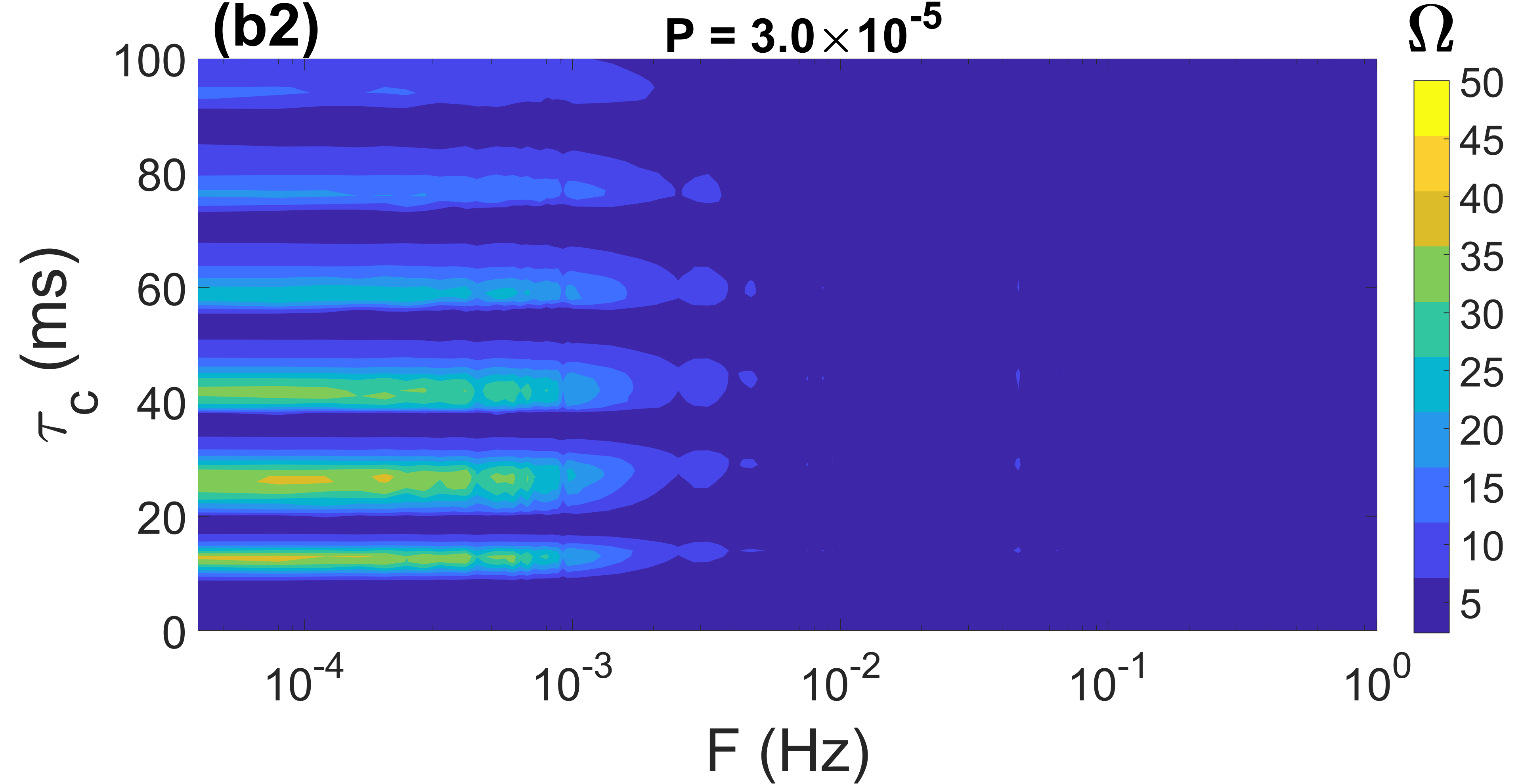}\includegraphics[width=5.5cm,height=3.41cm]{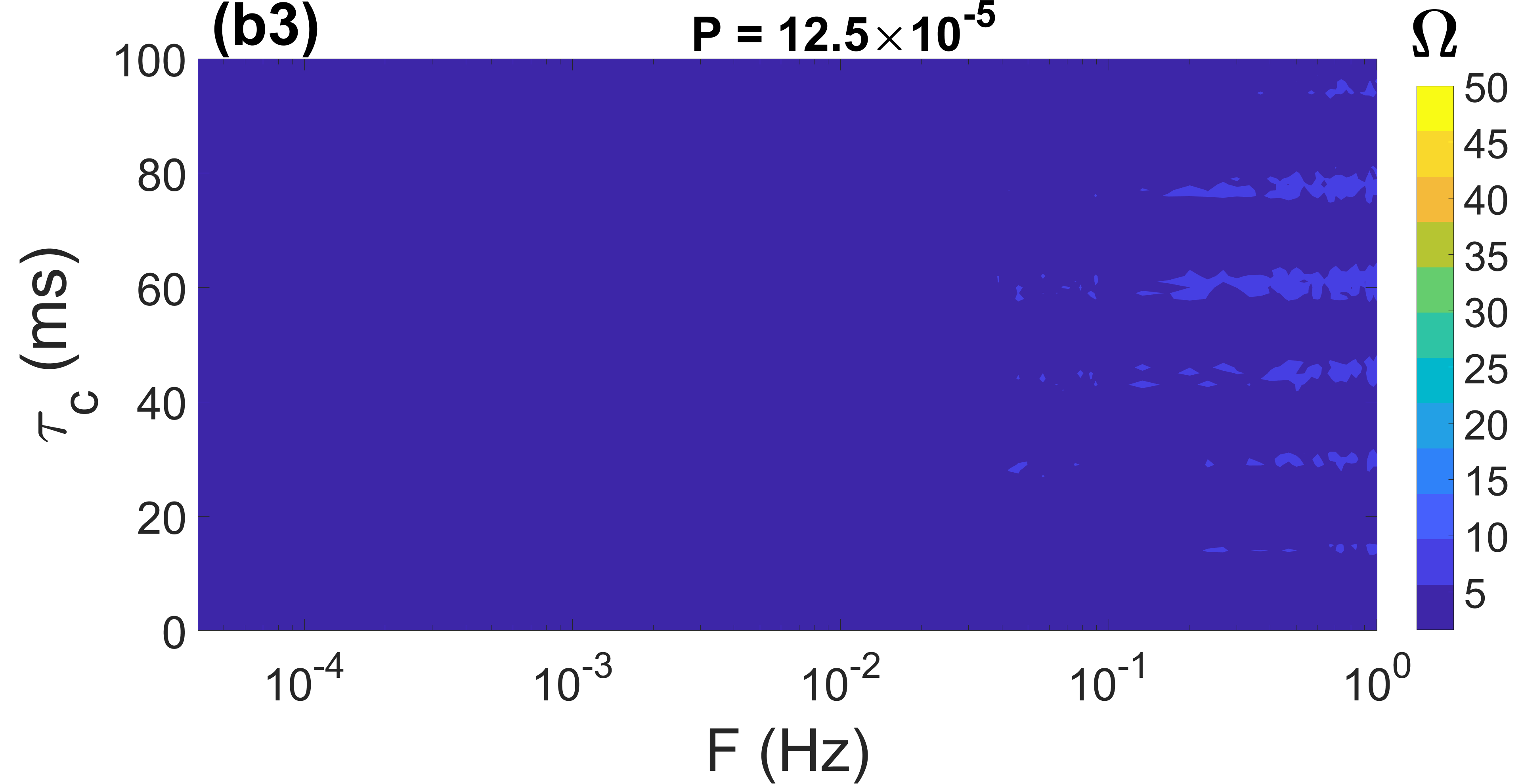}
\caption{ Inverse coefficient of variation $\Omega$ in the rewiring frequency $F$ and time delay $\tau_c$ parameter space in (a1)-(a3) a small-world network with $\beta=0.25$ and $\langle k\rangle=5$ at the indicated values of $P$; and (b1)-(b3) a random network with $\beta=1$ and $\langle k\rangle=5$ at the indicated values of the $P$.}
\label{fig:2}
\end{figure*}

In Figs. \ref{fig:2}(a1)-(a3), we show the contour plots of $\Omega$ against $F$, and the time delay $\tau_c$ in a small-world network at three values of $P$. In Fig. \ref{fig:2}(a1), it is seen that the smallest value of $P=0.1\times10^{-5}$ (i.e., when the weakening effect of STDP on the synaptic weights is the least pronounced) and a small value of $F<10^{-3}$ \marius{$\mathrm{Hz}$} (i.e., when the links switch slowly or remain static), the spiking coherence of the network is optimized only at some values of the time delay, i.e., at $\tau_c=\{13, 26, 42, 58, 76\}$.

In Fig. \ref{fig:2}(a2), we increase the weakening effect of STDP on the synaptic weights by increasing the value of the adjusting rate to $P=3.0\times10^{-5}$. It is shown that as $\tau_c$ increases, $\Omega$ also intermittently increases and decreases, indicating the presence of MCR. However, the intermittent peak values of $\Omega$ have smaller (compared to Fig. \ref{fig:2}(a1)) amplitudes. Furthermore, at the intermittent values of $\tau_c$ where $\Omega$ peaks, $F$ has a less significant (compared to Fig. \ref{fig:2}(a1)) effect on $\Omega$. 

In Figs. \ref{fig:2}(b1)-(b3), we show the contour plots of $\Omega$ against $F$ and $\tau_c$ in the random network at the three values of $P$. For $F<10^{-3}$ \marius{$\mathrm{Hz}$}, the results are qualitatively the same as in the small-world network in Figs. \ref{fig:2}(a1)-(a3).  However, irrespective of the value of $\tau_c$, when $F\ge10^{-3}$ \marius{$\mathrm{Hz}$}, there is a sudden drop in the degree of CR ($\Omega<5$), leading to the complete muting of MCR. 

In Figs. \ref{fig:2}(a3) and (b3), we further increase the weakening effect of the STDP rule on the synaptic weights by increasing the value of the adjusting rate to $P=12.5\times10^{-5}$ in the small-world and random network, respectively. Irrespective of $F$ and $\tau_c$,  we observe that in both types of networks, the degree of coherence is degraded significantly to very low levels, alongside the disappearance of MCR. It has been shown in \cite{yamakou2019control} that decreasing the coupling strength in a time-delayed FHN neural network increases the excitability of the network, making it more difficult and even impossible to achieve CR. In this viewpoint, the deterioration of coherence alongside the disappearance of MCR with increasing $P$ can be explained by the fact that larger values of $P$  (leading to a weakening of the time-delayed synaptic weight) induce a stronger degree of excitability, making it difficult to achieve a reasonably high degree of CR.  

\marius{
Next, we provide a theoretical explanation for the observation where the spiking time behavior of the neurons intermittently becomes ordered and less ordered, exhibiting MCR as $\tau_c$ increases. First, we recall that if a deterministic delayed differential equation (DDE) $\dot{x}= f(x(t),x(t-\tau_c))$, where $\tau_c$ is the time delay, possesses a solution $x(t)$ with period $\overline{\tau}$, then $x(t)$ also solves $\dot{x}= f(x(t),x(t-\tau_c-n\overline{\tau}))$, for all positive integers $n\in\mathbb{N}$. Our stochastic system of HH delayed differential equations could satisfy this property if it behaves like a deterministic delayed differential equation (DDE). This would be possible if our stochastic delayed HH equations could admit (quasi-) periodic solutions, i.e., (quasi-) deterministic spiking times. We know that (quasi-) deterministic spiking times can be achieved via CR, a phenomenon during which the inter-spike intervals in a time series are (almost) the same, leading to a period of neural activity $\overline{\tau}$ (i.e., the averaged inter-spike interval of the time series) which should be (approximately) equal or at least of the order of the distance between the periodic horizontal CR bands in Fig. \ref{fig:2}. The distance between the first four periodic horizontal CR bands in Fig. \ref{fig:2} is approximately $15.75$ $\mathrm{ms}$, while in Fig. \ref{fig:0}(b), the period of neural activity $\overline{\tau}$ at peak coherence is $\overline{\tau}=15.95$ $\mathrm{ms}$ --- they are pretty close, hence the phenomenon of MCR.
}

\marius{
Furthermore, we recall that the degree of CR always enhances (degrades) when we get closer (farther away) from the bifurcation thresholds \cite{pikovsky1997coherence,neiman1997coherence,hizanidis2008control,gu2011coherence}. The time delay in the nonlinear form of chemical synapses given in Eq.\eqref{eq:5a} has been shown to modulate the excitability of neural networks \cite{masoliver2017coherence,yamakou2019control}. 
Therefore, the observation where the peak values of $\Omega$ occurring at values $\tau_c=13$, 26, 42, 58, and 76 $\mathrm{ms}$ (corresponding to the horizontal bands of peak coherence in Fig. \ref{fig:2}) decreases $\tau_c$ increases, can be explained as follows: as the value of $\tau_c$ increases, the network gets farther away from the bifurcation thresholds (i.e., its degree of excitability is enhanced as $\tau_c$ increases --- confirmed by the simulations (not shown) where stronger noise intensities are required to induce spikes from the excitable regime when the time delay becomes larger), leading to a decrease in the degree of coherence provoked by the rarity of spikes. In this case, our stochastic delayed HH equations no longer behave like a deterministic DDE with the periodic property of the solutions stated above. Hence, the gradual and, eventually, the complete appearance of MCR as $\tau_c$ increases.}

\subsection{Combined effect of $F$, $\langle k \rangle$, and $P$}
\begin{figure*}
\centering
\includegraphics[width=5.5cm,height=3.41cm]{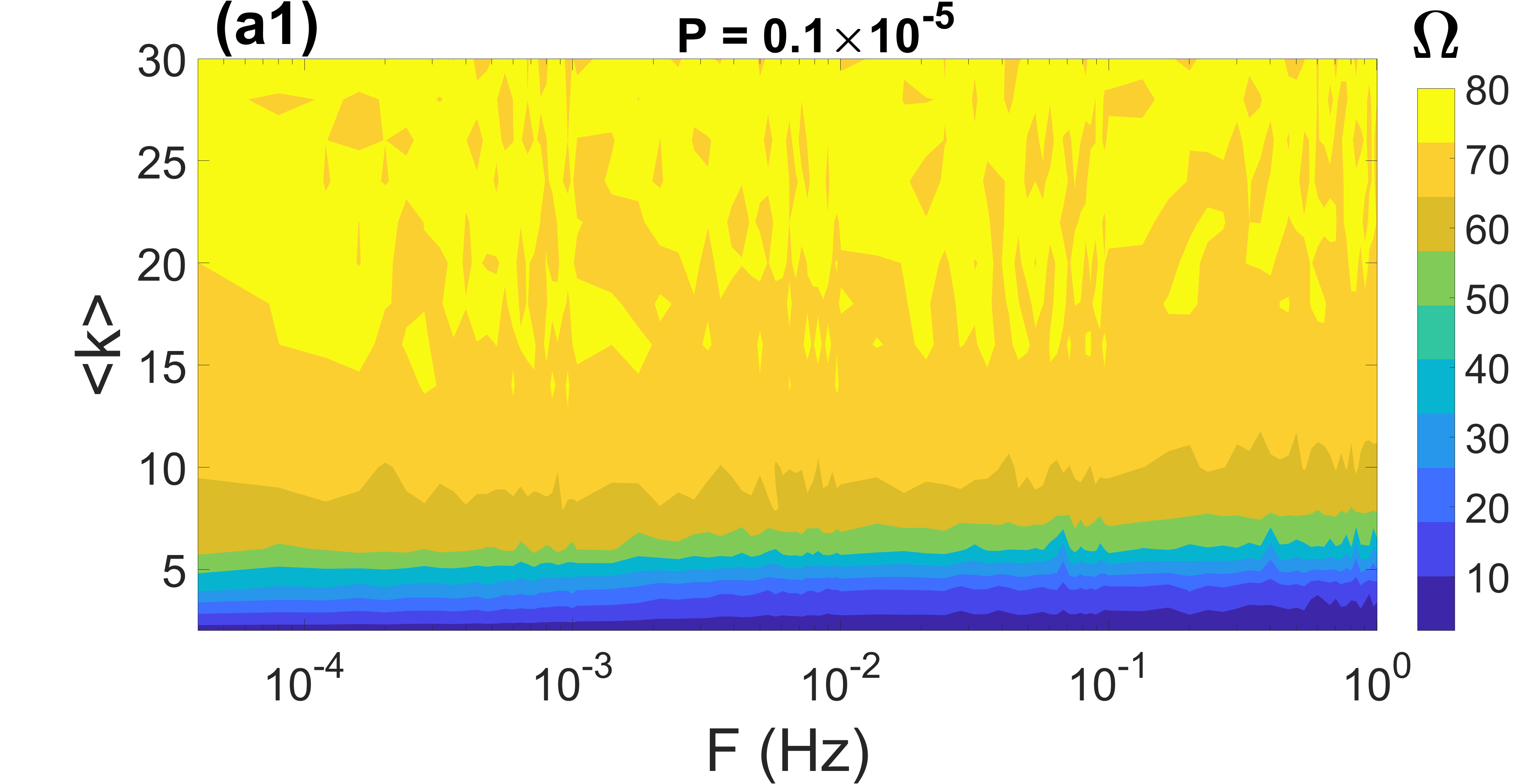}\includegraphics[width=5.5cm,height=3.41cm]{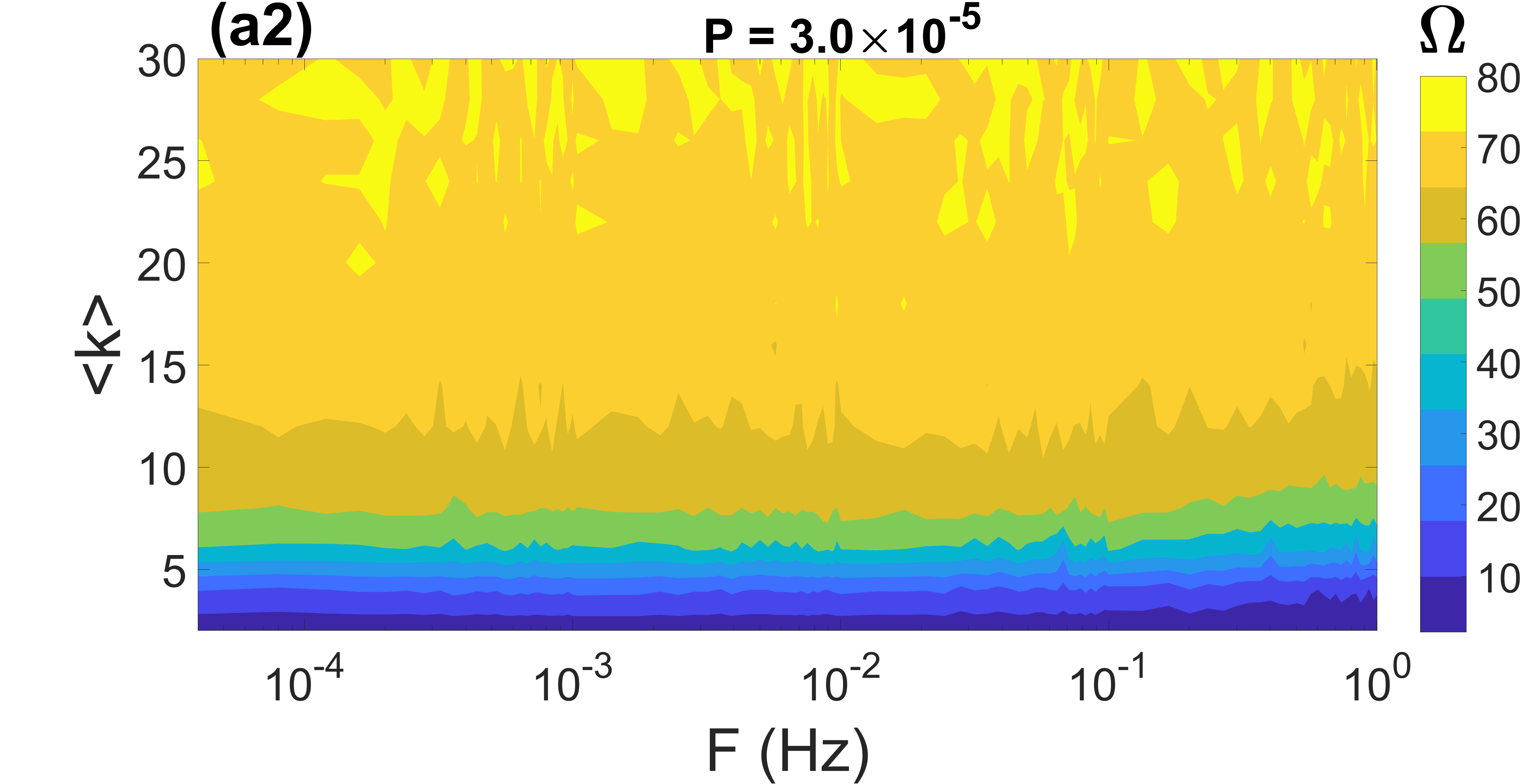}\includegraphics[width=5.5cm,height=3.41cm]{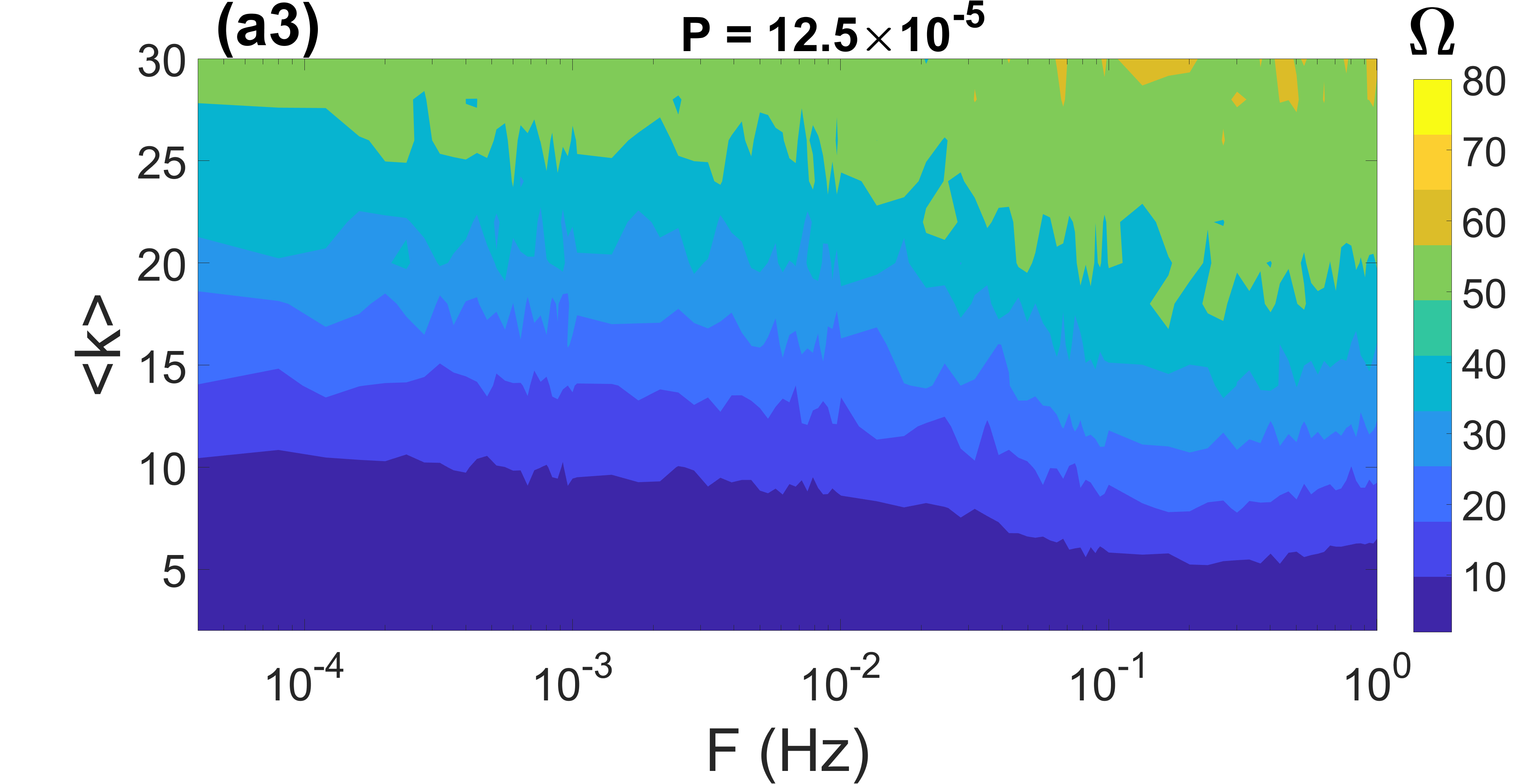}\\[2.0mm]
\includegraphics[width=5.5cm,height=3.41cm]{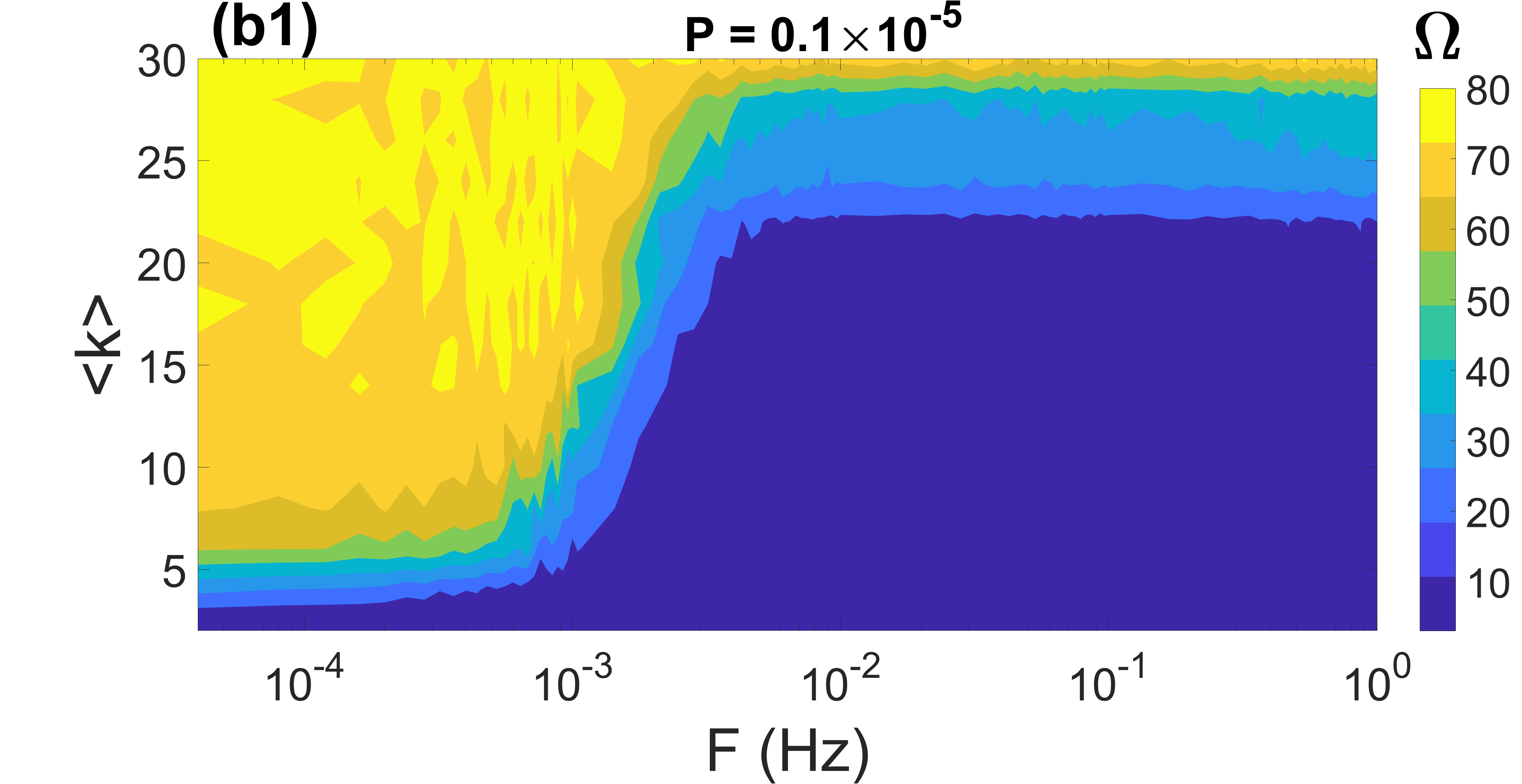}\includegraphics[width=5.5cm,height=3.41cm]{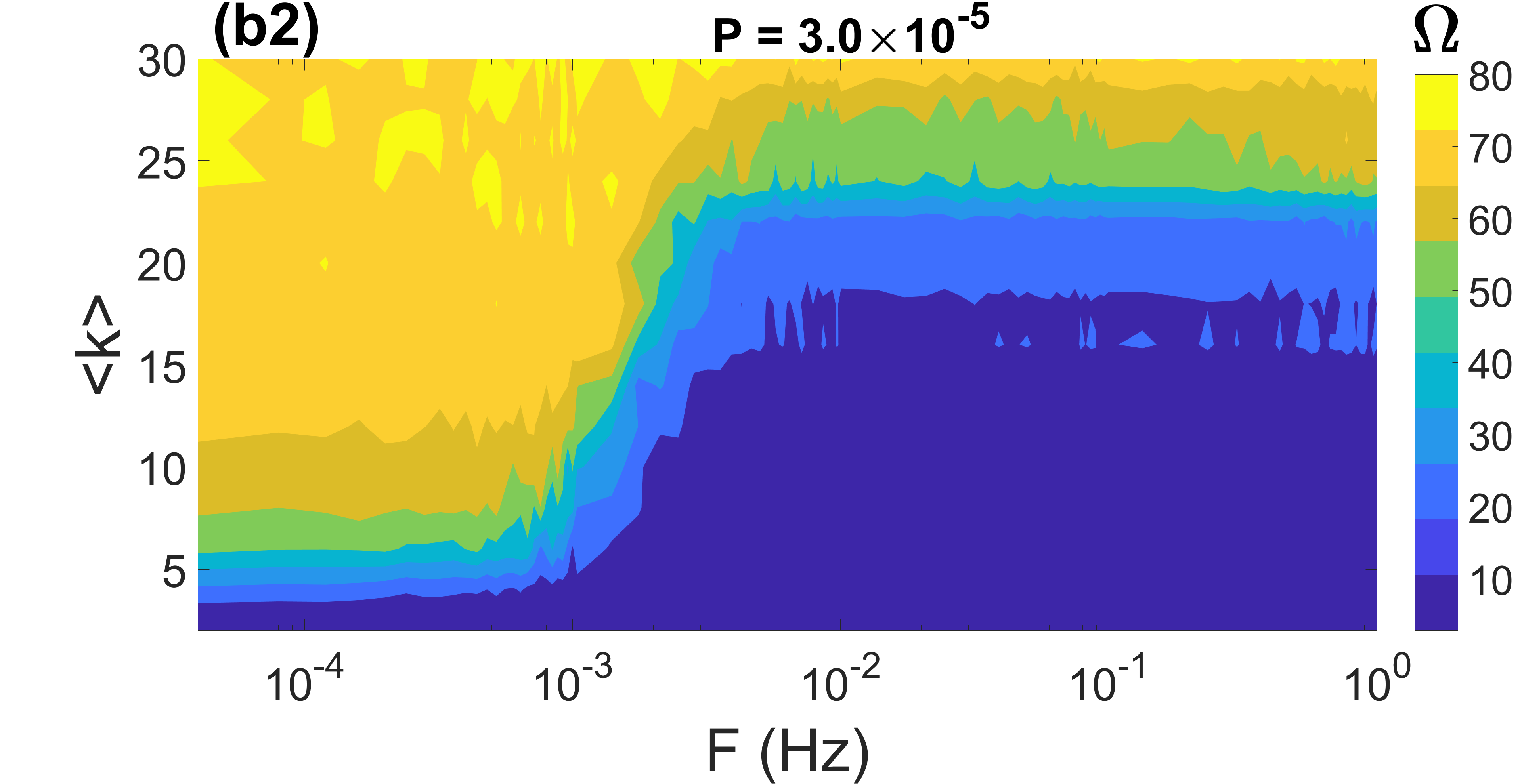}\includegraphics[width=5.5cm,height=3.41cm]{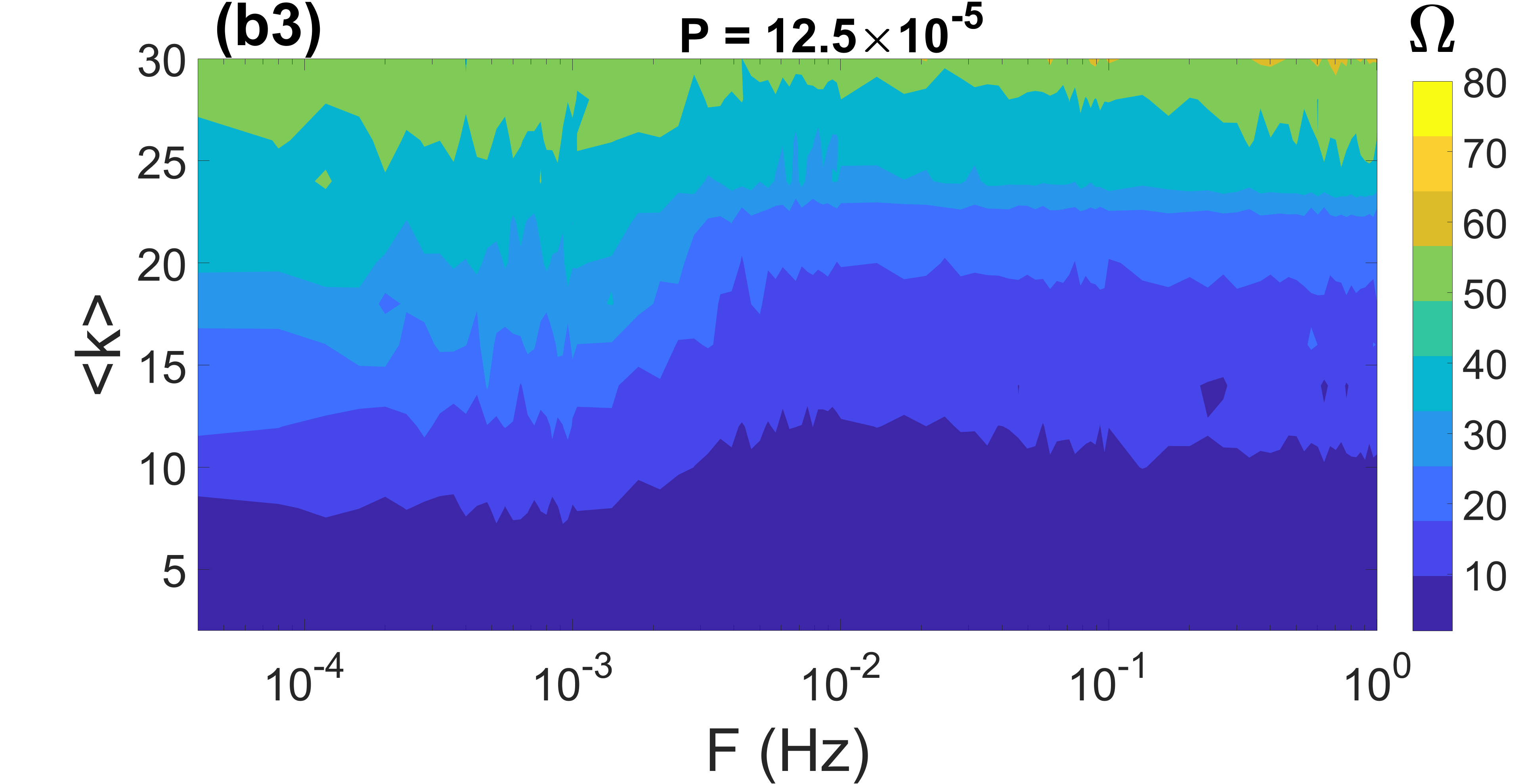}
\caption{ Inverse coefficient of variation $\Omega$ in the rewiring frequency $F$ and average degree $\langle k\rangle$ parameter space in (a1)-(a3) small-world network with $\beta=0.25$ and $\tau_c=13.0$ \marius{$\mathrm{ms}$} at the indicated values of $P$; and (b1)-(b3) a random network with $\beta=1$ and $\tau_c=13.0$ \marius{$\mathrm{ms}$} at the indicated values of $P$.}
\label{fig:3}
\end{figure*}

In Figs. \ref{fig:3}(a1)-(a3), we show the contour plots of $\Omega$ against $F$, and the average degree $\langle k \rangle$  in a small-world network ($\beta=0.25$) at the same three previous values of $P$, with the time delay at $\tau_c=13.0$ \marius{$\mathrm{ms}$}, i.e., a value of $\tau_c$, from Fig. \ref{fig:2}, at which we have the highest degree of CR.
In each of these figures, we observe that irrespective of the value of $F$, larger values of $\langle k \rangle$ induce a higher degree of CR. \marius{This behavior can be explained by the fact that with higher values of $\langle k \rangle$, the network becomes denser, leading to more interaction between the neurons in the network. This can, in turn, facilitate synchronization in the network where \R{less coherent neurons (with low $\Omega$ values) synchronize more coherent neurons (with high $\Omega$ values)}. This has an overall effect of increasing the average $\Omega$ of the network, thus enhancing CR. }

\marius{On the other hand, as the network becomes
less dense (i.e., with smaller values of $\langle k \rangle$), all the neurons in the network can no longer so easily synchronize (in particular, of course, those which are not connected); hence, the averaged $\Omega$ of the network is calculated with low $\Omega$ values (of less coherent neurons which cannot easily synchronize due to sparsity of the network) and high $\Omega$ values (of more coherent neurons). This has the overall effect of shifting the
averaged $\Omega$ to lower values, thus deteriorating CR.} Moreover, in Figs. \ref{fig:3}(a1)-(a3), we observe that smaller values of $P$ increase the degree of CR. This is explained by the fact that smaller $P$ strengthens the synaptic weights between neurons (see Figs. \ref{fig:1}(a2) and (b2)) and hence improves their synchronization (see Figs. \ref{fig:1}(a3) and (b3)), which lead to a better degree of CR.  

In Figs. \ref{fig:3}(b1)-(b3), we show the contour plots of $\Omega$ against $F$, and the average degree $\langle k \rangle$  in the random network ($\beta=1$) at the same three previous values of $P$, with the time delay at $\tau_c=13.0$ \marius{$\mathrm{ms}$}. First, we observe that in these figures, increasing $\langle k \rangle$ increases the degree of CR for the same reasons we gave for the case of the small-world network. However, in contrast to the small-world network where only $\langle k \rangle$ significantly affects the degree of CR, both $\langle k \rangle$ and $F$ significantly affect the degree of CR in the random network. In particular, we observe that in Figs. \ref{fig:3}(b1) and (b2), for $F\in[0,10^{-3}]$ \marius{$\mathrm{Hz}$} we have a higher degree of CR as $\langle k \rangle$ increases. However, when $F>10^{-3}$ \marius{$\mathrm{Hz}$}, the high degree of CR deteriorates significantly—comparing Figs. \ref{fig:3}(b1)-(b3), we see that increasing the value of $P$ (i.e., weakening the synaptic weights and hence poorer synchronization) leads to a lower degree of CR.

\subsection{Combined effect of $F$, $\beta$, and $P$}
\marius{All the results presented for the small-world network in Figs. \ref{fig:0}-\ref{fig:3} are obtained with a rewiring probability of $\beta=0.25$. In this subsection, for the sake of completeness, we explore the variation in the degree of CR when the small-world networks are generated with different values of the rewiring probability.}

\begin{figure*}
\centering
\includegraphics[width=5.5cm,height=3.41cm]{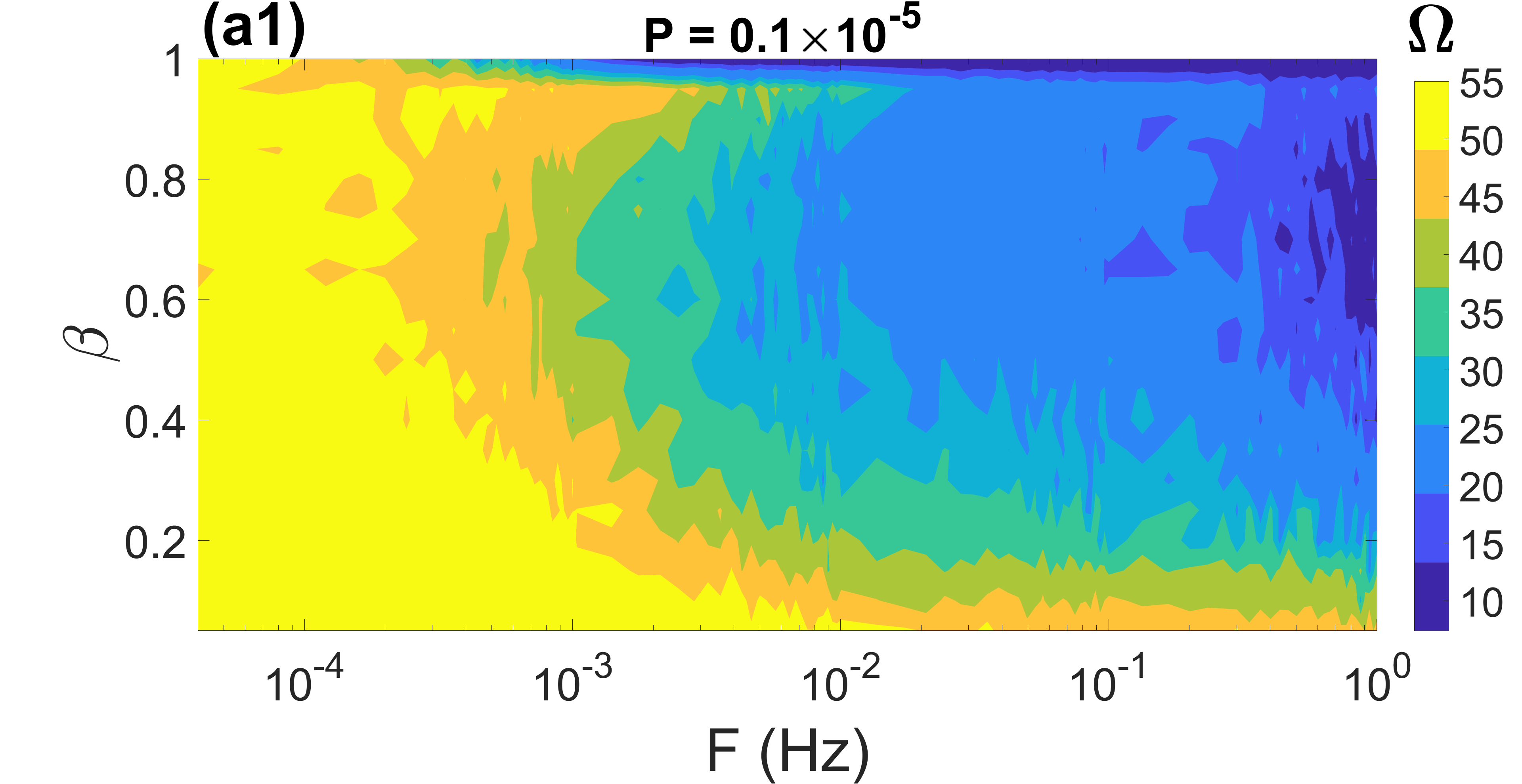}\includegraphics[width=5.5cm,height=3.41cm]{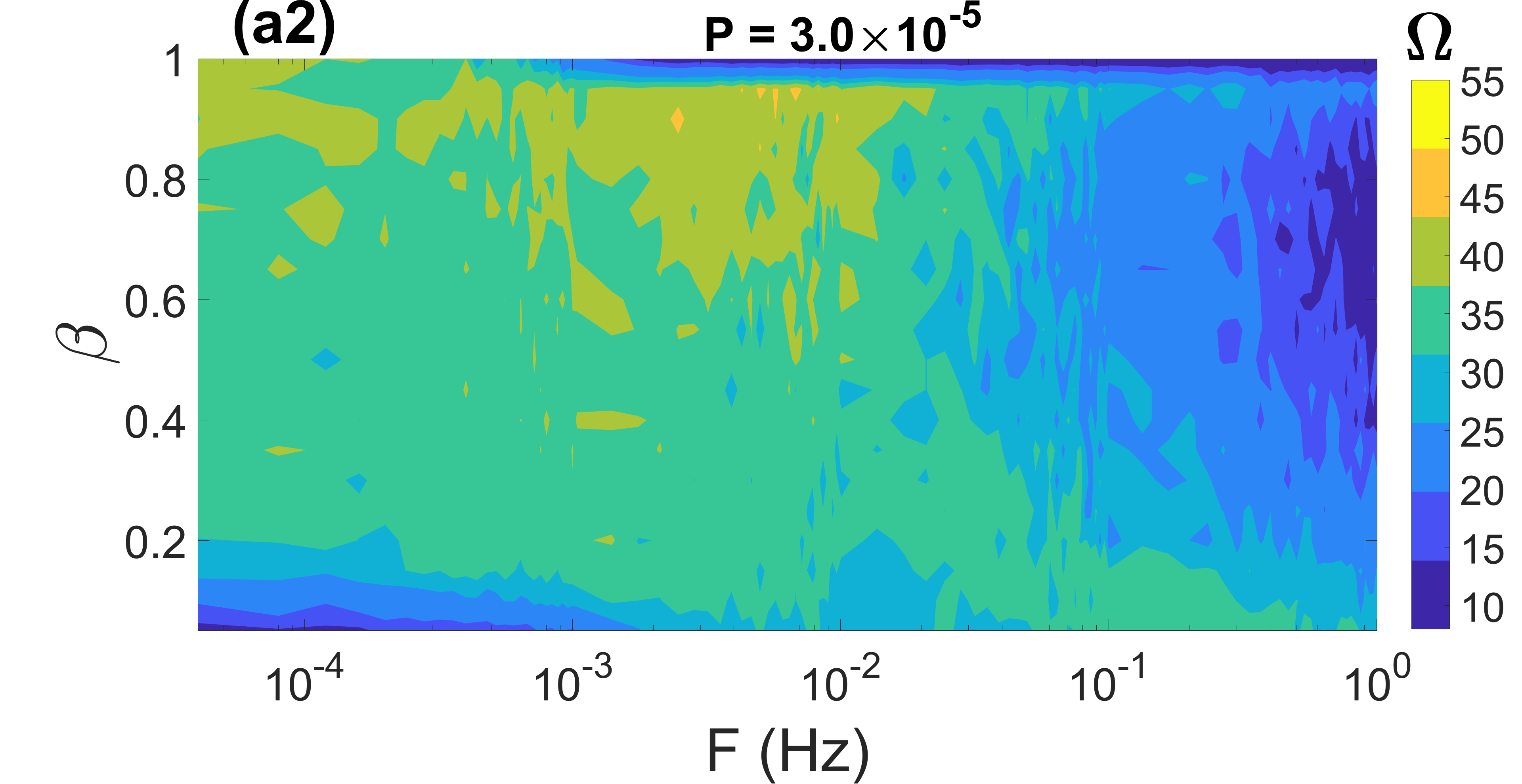}\includegraphics[width=5.5cm,height=3.41cm]{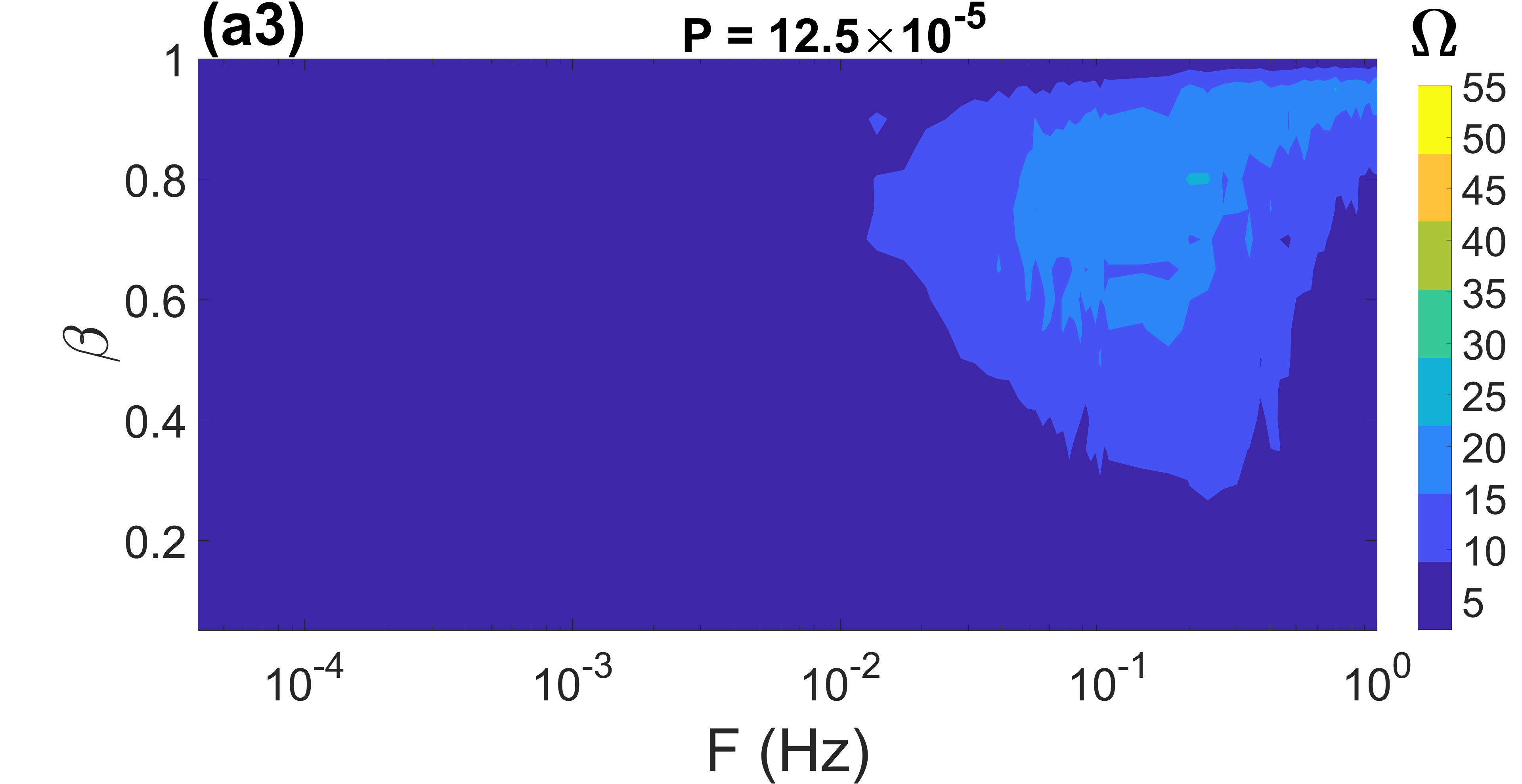}\\[2.0mm]\includegraphics[width=5.5cm,height=3.41cm]{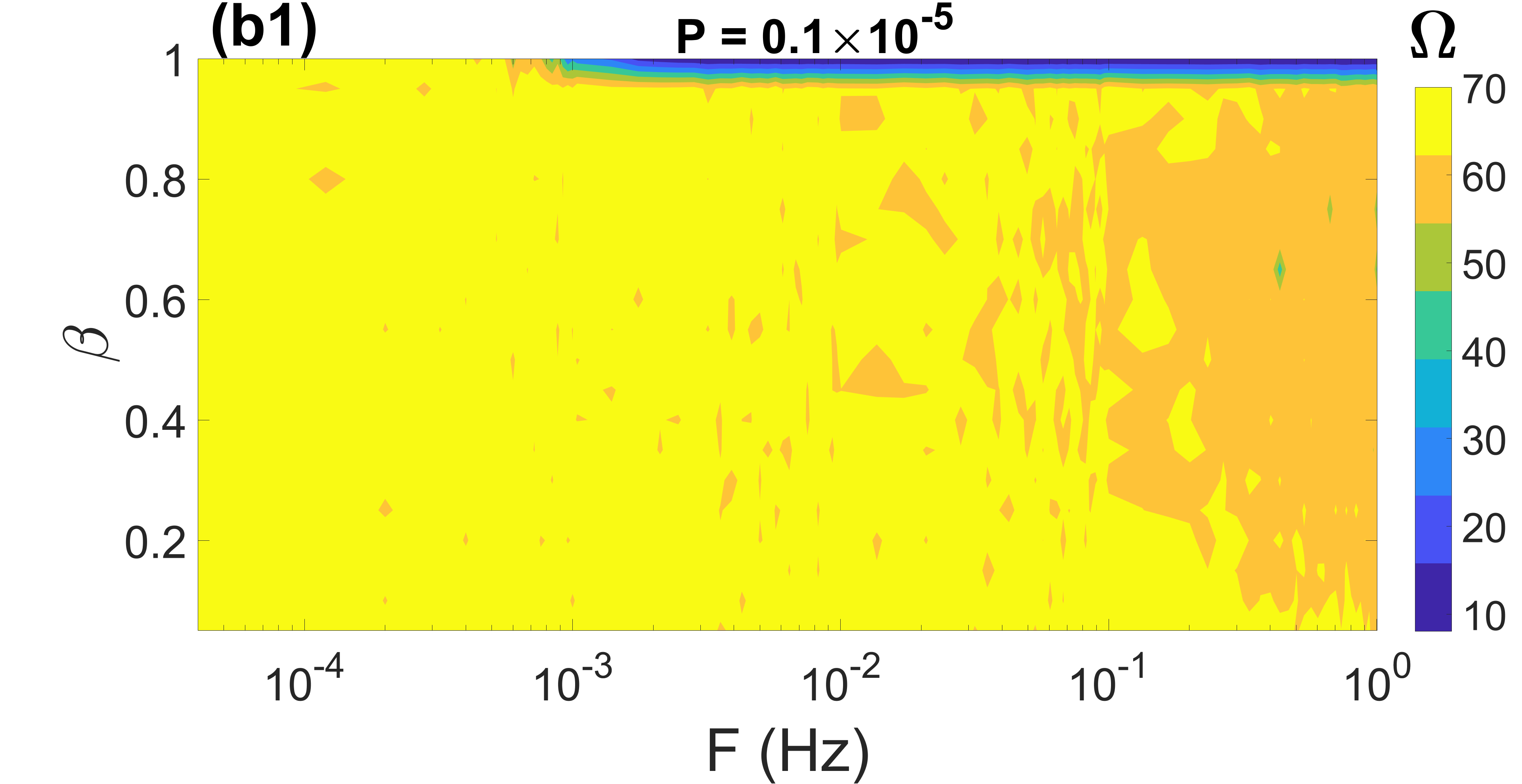}
\includegraphics[width=5.5cm,height=3.41cm]{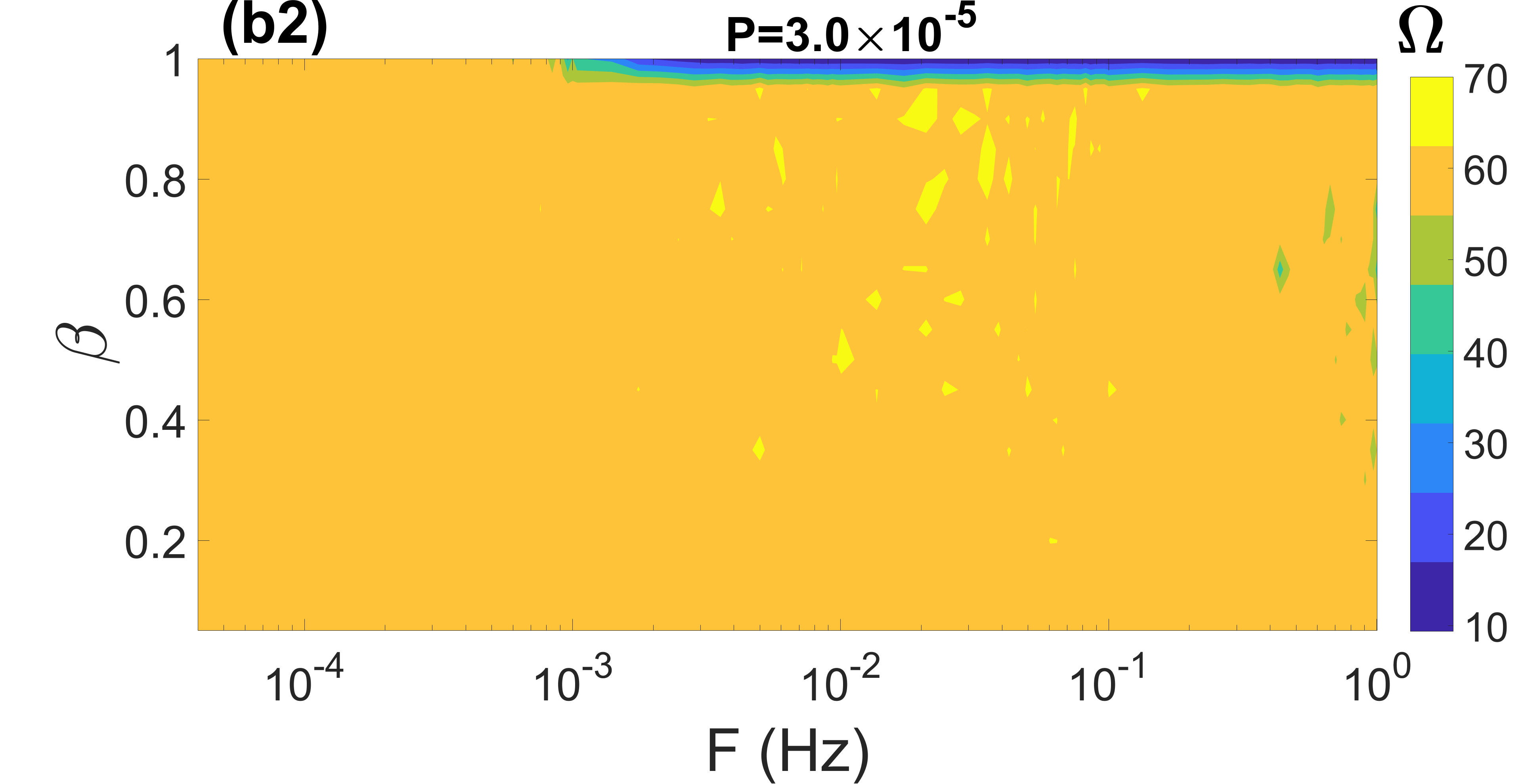}\includegraphics[width=5.5cm,height=3.41cm]{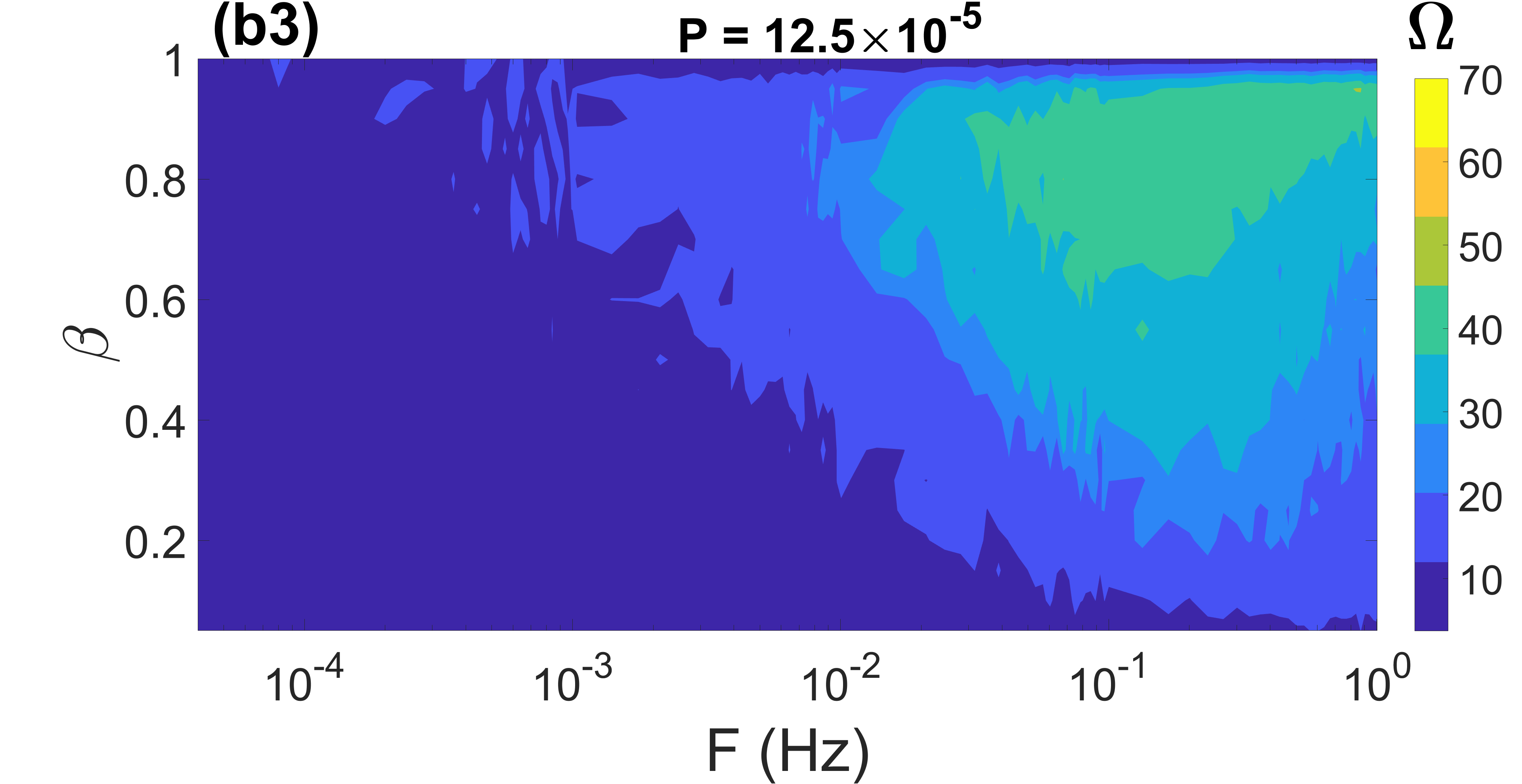}
\caption{Inverse coefficient of variation $\Omega$ in the rewiring frequency $F$ and rewiring probability $\beta$ parameter space in (a1)-(a3) small-world ($0.05\leq\beta<1$) and random ($\beta=1$) networks with $\tau_c=13.0$ \marius{$\mathrm{ms}$} and $\langle k\rangle=5$ at the indicated values of $P$; and (b1)-(b3) small-world ($0.05\leq\beta<1$) and random ($\beta=1$) with $\tau_c=13.0$ \marius{$\mathrm{ms}$} and $\langle k\rangle=10$ at the indicated values of $P$.}
\label{fig:4}
\end{figure*}

In Figs. \ref{fig:4}(a1)-(a3), we show the contour plots of $\Omega$ against $F$, and the rewiring probability $\beta\in[0.05,1]$ (which increases with the number of random short cuts in the small-world network) at the same three previous values of $P$, a time delay of $\tau_c=13.0$ \marius{$\mathrm{ms}$}, and a relatively low average degree of $\langle k \rangle=5$. 

In Fig. \ref{fig:4}(a1), with a very low value of $P$ (which strengthens the synaptic weights of the networks), both the small-world ($\beta\in[0.2,1)$) and random ($\beta=1$) networks have a very high degree of CR if the rate at which the networks rewire is relatively low, i.e., $F\in[0,10^{-3}]$ \marius{$\mathrm{Hz}$}. We also see that if the number of random shortcuts in the small-world network is low (i.e., $\beta<0.2$), then all the values of $F$ do not significantly affect the high degree of CR. But this high degree of CR deteriorates significantly for $\beta>0.2$ and for $F>10^{-3}$ \marius{$\mathrm{Hz}$}.

In Fig. \ref{fig:4}(a2), with an intermediate value of $P$, the behavior of CR in the random network (i.e., when $\beta=1$) remains qualitatively the same when compared to Fig. \ref{fig:4}(a1). Except that the degree of CR has decreased because of the weakening of the synaptic weights induced by the larger value of $P$. However, in the small-world networks, i.e., when $\beta\in[0.05,1)$, the behavior of CR has changed qualitatively and quantitatively. In particular, we observe that for \marius{$\beta\approx0.05$}, higher rewiring frequencies $F>10^{-3}$ \marius{$\mathrm{Hz}$} are required to enhance the degree of CR. This behavior contrasts with that in Fig. \ref{fig:4}(a1), where higher $F$ gradually deteriorates the degree of CR.

In Fig. \ref{fig:4}(a3), we have a relatively large value of $P$. 
For the random network (i.e., $\beta=1$), all rewiring frequencies deteriorate the degree of CR, in contrast to Figs. \ref{fig:4}(a1) and (a2)  with low and intermediate values of $P$, where the random network will exhibit a high degree of CR at small values of $F(<10^{-3})$ and a low degree of CR at higher values of $F(>10^{-3})$. For the small-world networks (i.e., $\beta\in[0.05,1)$), all the values of $\beta$ and the $F$ deteriorate the degree of CR, except for relatively higher values $\beta\in[0.55,0.85)$ and $F(>10^{-1})$, where the intermediate degree of CR is maintained from the two previous cases with low and intermediate values of $P$.

In Figs. \ref{fig:4}(b1)-(b3), we show the contour plots of $\Omega$ against $F$, and the rewiring probability $\beta\in[0.05,1]$ (which increases with the number of random short cuts in the small-world network) at the same three previous values of $P$, a time delay of $\tau_c=13.0$ \marius{$\mathrm{ms}$}, and a higher (compared to Figs. \ref{fig:4}(a1)-(a3)) average degree of $\langle k \rangle=10$.  First, when we compare Figs. \ref{fig:4}(b1)-(b3), we see that, again, larger values of $P$ deteriorate the degree of CR. Secondly, for the random networks (i.e., when $\beta=1$), a combination of slowly varying connections (i.e., when $F<10^{-3}$ \marius{$\mathrm{Hz}$}) with low or intermediate values of $P$ enhances the degree of CR. However, at higher values of $P$, even these small rewiring frequencies ($F<10^{-3}$ \marius{$\mathrm{Hz}$}) cannot enhance the low degree of CR. For the small-world networks, i.e., when $\beta\in[0.05,1)$, the relatively higher value of the average degree $\langle k \rangle=10$ promotes more interaction in the denser network, which favors synchronization in addition to the stronger synaptic weights induced by the low and intermediate values of $P$ in Figs. \ref{fig:4}(b1) and (b2). 

In Fig. \ref{fig:4}(b3), the relatively larger value of $P$ has deteriorated the degree of CR. However, higher rewiring frequencies ($F>10^{-2}$ \marius{$\mathrm{Hz}$}) induce a more enhanced degree of CR in contrast to  Fig. \ref{fig:4}(b1), where the highest degree of CR is achieved with relatively lower rewiring frequencies.
Thus, Fig. \ref{fig:4} indicates that increasing the number of random shortcuts (which increases with $\beta\in[0.05,1)$ in a small-world network), lower rewiring frequencies $(F<10^{-1}$ \marius{$\mathrm{Hz}$}) outperform the higher ones $(F>10^{-1}$ \marius{$\mathrm{Hz}$}) when the STDP parameter is small, i.e., $P=0.1\times10^{-5}$. But when $P$ becomes larger (e.g., $P=12.5\times10^{-5}$), higher rewiring frequencies $(F>10^{-1}$ \marius{$\mathrm{Hz}$}) outperform the lower ones ($F<10^{-1}$ \marius{$\mathrm{Hz}$}) as $\beta\in[0.05,1)$ increases. 

\section{Summary and conclusions}\label{Sec. V}
In summary, we have numerically investigated the phenomenon of CR in adaptive small-world and random neural networks driven by STDP and HSP. It is found that the degree of CR strongly depends on the adjusting rate parameter $P$, which controls STDP, and the characteristic rewiring frequency parameter $F$, which controls HSP. Decreasing $P$ (which increases the \R{strengthening} effect of STDP on the synaptic weights) and decreasing $F$ (which slows down the swapping rate of synapses between neurons) leads to a higher degree of CR in both the small-world (depending on the value of rewiring probability $\beta$) and random networks. It is found that the synaptic time delays $\tau_c$ can induce multiple CR (MCR) in both small-world and random networks, with MCR becoming more pronounced at smaller values of both $P$ and $F$. Within the $P-F$ parameter regime in which MCR occurs, increasing the time delay reduces the peak values of the inverse coefficient of variation and, thus, the degree of CR. It is also found that irrespective of the rewiring frequency $F$, the degree of CR increases when the average degree $\langle k \rangle$ in small-world increases. However, for a given average degree and rewiring frequency, higher values of the adjusting rate parameter $P$ turn to deteriorate the degree of CR.

On the other hand, for random networks, the increase in the degree of CR with the increase in the average degree $\langle k \rangle$ depends on the rewiring frequency. With higher rewiring frequencies ($F>10^{-3}$ \marius{$\mathrm{Hz}$}), a larger average degree is required to enhance the degree of CR in the random network. Furthermore, it is also found that while large values of $F$ ($>10^{-1}$ \marius{$\mathrm{Hz}$}) can either deteriorate (when $P$ is small) or enhance (when $P$ are relatively large), the degree of CR in small-world networks (with $\beta\in[0.05,1)$ ), in random networks (when $\beta=1$) they can only deteriorate the degree of CR, regardless of the value of $P$. 

It is worth noting that the results presented in this paper may be sensitive to the choice of rewiring strategies used to maintain the small-worldness and the complete randomness of the time-varying small-world and random networks, respectively. At this point, nevertheless, our results have the implication that inherent background noise, the prevalent spike-timing-dependent plasticity, and homeostatic structural plasticity can jointly play a significant constructive role in enhancing the time precision of firing, when the right amount of time delays, average degree, and randomness are introduced in the neural systems. 

\marius{
There is strong experimental evidence that acetylcholine, monoamines, and other signaling molecules can control STDP \cite{brzosko2019neuromodulation}. Also, the control of synapses in the brain has become more accessible via drugs that affect neurotransmitters \cite{pardridge2012drug}. Thus, our results could guide the control of synapses and STDP for optimal neural information processing via CR in electrophysiological experiments. Neuromorphic engineering is an active and rapidly developing field where engineers designed bio-inspired artificial neural circuits to process information differently to perform specific computational tasks \cite{panzeri2022constraints,eberhardt1989vlsi}. Thus, our results could find applications in the design of \textit{ad hoc} artificial neural circuits engineered to use CR to optimize signal processing.}


\begin{widetext}
\section*{appendix}\label{algo}
\begin{algorithm}[H]
\tcc{ 
$X_i(t)$ $=\{ V_i(t),\:m_i(t),\:h_i(t),\:n_i(t)$: variables of coupled SDEs in Eqs.\eqref{eq:1} and \eqref{eq:2}\;
$t$: time\;
$T$: total integration time\;
$N$: network size\;
$M$: number of realizations\;
$F$: rewiring frequency of synapses\;
$F_{max}$: maximal rewiring frequency\;
$P$: STDP control parameter\;
$P_{min}$: min of $P$\;
$P_{max}$: max of $P$\;
$\ell_{ij}(t)$: adjacency  matrix\;
$g_{ij}(t)$: synaptic weights\; 
$g_m$: average of synaptic weights over $t$ and $N$ of the $m^{th}$ realization\;
$\beta$: rewiring probability in Watts-Strogatz algorithm\;
$t^n_i$ : $n^{th}$ spike time of the $i^{th}$ neuron\;
$\omega_m$: inverse  coefficient of variation of the $m^{th}$ realization\;
$r_m$: order parameter of the  $m^{th}$ realization\;
$\Omega$: average inverse coefficient of variation over $M$\;
$R$: average order parameter over $M$\;
$G$: average of synaptic weights over $M$\;
}	
\KwInput{$T$, $N$, $M$, $F$, $P$, $\beta$}
\KwOutput{$\Omega, G, R$}
$P \gets P_{min}$ \tcp*{Initialize the adjusting rate parameter}
\While{$ P \leq P_{max}$ }  { 
$F\leftarrow 0$ \tcp*{Initialize the rewiring frequency}
	\While{$ F \leq F_{max}$ } {
		\For{$m \in  1,2,\dots,M$}{
			Init $X_i(t)\:,\:\ell_{ij}(t)$ \tcp*{Random initial conditions of SDEs and initial SW or RND network adjacency matrix}
			\For{$t \in 0,\dots,T $}{
				Integrate network of SDEs in Eqs.\eqref{eq:1} and \eqref{eq:2}\tcp*{Using the Euler-Maruyama method}
				Record the current voltage spike times $t^{n}_i$ from $V_i(t)$\tcp*{Times $t$ at which $V_i(t)\ge V_{\mathrm{th}}=0.0$}
				\If{$\Delta t_{ij}:=t_i - t_j > 0$}  {
					$\Delta M\gets P\exp{(-\lvert\Delta t_{ij}\rvert/\tau_{p})}$ 
			\tcp*{$t_i$\:,\:$t_j$: nearest-spike times of post ($i$) \& pre ($j$) neuron} 
			}
		
				\If{$\Delta t_{ij} <0$} {
				$\Delta M \gets  -1.05 P\exp{(-\lvert\Delta t_{ij}\rvert/\tau_{d})}$ 
			}
		
			\If{$\Delta t_{ij} = 0$} {
				$\Delta M \gets  0$ 
			}
				$g_{ij}(t) \gets g_{ij}(t) + g_{ij}(t)\Delta M$\tcp*{update synaptic weights}
					$\ell_{ij}(t) \gets \widetilde{\ell_{ij}}(t) $ \tcp*{Update the adjacency matrix with  $\widetilde{\ell_{ij}}(t)$ obtained by randomly rewiring $\ell_{ij}(t)$ with 
     frequency $F$ according to the small-world ($\beta\in(0,1)$) or random ($\beta=1$) network rewiring strategy}
					}
			$\overline{\langle \mathrm{\tau} \rangle}\gets \big \langle  \big \langle t_i^{n+1} - t_i^n \big\rangle_{t}\big\rangle_{{N}}$,\:
			$\overline{\langle \mathrm{\tau}^2 \rangle} \gets \big \langle  \big \langle (t_i^{n+1} - t_i^n)^2 \big\rangle_{t}\big\rangle_{{N}}$ \tcp*{Compute mean \& mean squared ISI (on $t$ and $N$)}

			$\omega_m \gets  \displaystyle{\frac{\overline{\langle \mathrm{\tau} \rangle}}{\sqrt{\overline{\langle \mathrm{\tau}^2 \rangle} - \overline{\langle \mathrm{\tau} \rangle^2}}}}$\:,
			$g_m \gets \displaystyle{\bigg \langle N^{-2}\sum\limits_{i=1}^{N}\sum\limits_{j=1}^{N}g_{ij}(t)\bigg \rangle_t}$\;
			$r_m \gets \displaystyle{\bigg \langle\bigg|N^{-1}\sum\limits_{i=1}^N\exp{\Big[j\Big(2\pi \ell + 2\pi(t-t_{_{i}}^{n})/(t_{_{i}}^{n+1}-t_{_{i}}^{n})\Big)\Big]}\bigg|\bigg \rangle_t}$ \tcp*{where $j=\sqrt{-1}$}
			
			Add $\omega_m$ to $\omega$\;
			Add $g_m$ to $g$\;
			Add $r_m$ to $r$\;	
		}
		$\Omega \gets \omega/M$ \tcp*{Compute averages over the $M$ realizations}
		$G \gets g/M$ \tcp*{Compute averages over the $M$ realizations}
		$R \gets r/M$ \tcp*{Compute averages over the $M$ realizations}
		$ F \gets F + \Delta F$ \tcp*{Increment the HSP control parameter}
	} 
$P \gets P + \Delta P$  \tcp*{Increment the STDP control parameter}
\caption{Flow of control in the simulations}
}
\end{algorithm}
\end{widetext}

\begin{acknowledgments}
MEY acknowledges support from the Deutsche Forschungsgemeinschaft (DFG, German Research Foundation) -- Project No. 456989199.
\end{acknowledgments}



\providecommand{\noopsort}[1]{}\providecommand{\singleletter}[1]{#1}%
%


\end{document}